\documentstyle[epsf,12pt,subfigure]{article}

\title{Receptive Field and Feature Map Formation in the Primary Visual Cortex via Hebbian Learning with Inhibitory Feedback}
\author{
	Ted Hesselroth \and Klaus Schulten\\
	Department of Physics and  Beckman Institute\\
	University of Illinois\\
	Urbana, Illinois 61801\\[1ex]
	}

\begin{document}
\maketitle
\eject
{\em Abstract.}

{\small A linear neural network is proposed for mamalian vision system in which backward connections from the primary visual cortex (V1) to  the lateral geniculate nucleus play a key role. The backward connections control the flow of information from the LGN to V1 in such a way as to maximize the rate of transfer of  information from the LGN to V1. The application of hebbian learning to the forward and backward connections causes the formation of receptive fields which are sensitive to edges, bars, and spatial frequencies of preferred orientations. Receptive field types in V1 are shown to depend on the density of the afferent connections in the LGN. Orientational preferences are organized in the primary visual cortex by the application of lateral interactions during the learning phase. Change in the size of the eye between the immature and mature animal is shown be an important factor in the development of V1 organization. The orgainization of the mature network is compared to that found in the macaque monkey by several analytical tests.}
\section{Introduction}

The companion paper to this one may be found at http://arxiv.org/abs/q-bio.NC/0505010.

 One of the most interesting aspects of the the various areas of sensory processing in the brain is the presence of feature maps\cite{MOUN78A,BLAS86,RITT91A}. It has been found that individual neurons are generally responsive to some specific sensory feature of the input received, and that the neurons are organized in a continuous way in the cortical sheet in accordance with the features. In the somatosensory cortex, there is a map of the body; sensations from particular areas of the skin evoke activities at corresponding points in this area. In auditory processing the map is in terms of the frequency of a sound and its direction relative to the observer.

It has been found that neurons in the primary visual cortex are organized according to their feature detecting properties\cite{MALS73,BLAS91B,CHAP91A}. In particular, the orientational selectivity varies in a continuous way across the cortex to form a feature map. Neurons in the feature map encode for position as well as orientation, and this is reflected in the map's being retinotopic, that is, showing a one-to-one correspondence between positions in the two-dimensional cortex and positions in the visual field.  Because they are so large, the receptive fields of neighboring V1 neurons overlap considerably. In fact, since by continuity the neurons also have similar orientational selectivity, neighboring neurons encode almost exactly the same information. Such a redundancy in the V1 representation of the image is addressed in the model presented\cite{PECE92A,DAUG85A}.
 
For simple cells in the primary visual cortex (also called the V1 layer), feature selectivity is accomplished through spatial summation of the inputs. The receptive field, or area of response of a neuron in the visual field, is generally arranged as alternating excitatory and inhibitory regions, elongated in a particular direction\cite{HUBE62}. In this way, the neuron responds most strongly to  edges of a particular orientation or image components of specific spatial frequency and orientation. The sharpness of the tuning for orientation is of a width of about 20-40 degrees.  In the foveal region, the receptive fields are about one-quarter degree across\cite{HUBE74B}. The typical foveal simple cell's receptive field is a bar-detector, a thin line two minutes of arc across straddled by regions of opposite response characteristic, though there are also edge detector receptive fields  containing only two regions, excitatory and inhibitory.

Various models of receptive fields and feature maps have been 
proposed. In the following are summarized the three major types of 
models:

\subsection{Models of the Receptive Field Properties Only}

These models seek primarily to explain the basis of orientational 
selectivity in individual V1 neurons. Orientational preference has often 
been thought to come about through an alignment of geniculate 
inputs\cite{HUBE62}, and this idea has received some direct 
experimental confirmation\cite{CHAP91A}. Most of the theories are 
consistent with this and involve using Hebb's Rule to develop feature-
detecting properties from correlations in input data, as in the present 
model, arriving at connection strength arrangements which resemble 
the receptive fields of V1 neurons. A partial list is refs. 
\cite{SANG89,YUIL89A,KAMM88,HANC92,WORG90,WORG92B,WORG93,HUMM79,MARR82A,RUBN90,RUBN90A,KAMI93,LIZH93,NASS75,COOP79,WATS90,MCIL91},
not including related models of receptive fields in the 
retina.

\subsection{Low-Dimensional Models of V1 Self-Organization}

In low-dimensional models, mathematical functions or algorithms are used with outputs that imitate the functional organization of the visual cortex, resulting in mappings of retinotopy, ocular dominance, and orientational selectivity onto V1 \cite{ROJE90,ROJE90A,SWIN82,SWIN91,SWIN92,DURB90,GOOD90A,TAKE79A,OBER90A,OBER90C,OBY-91B,OBY-92C}. The models are called low-dimensional because some high-dimensional quantities, such  as sets of connection strengths, are not included explicitly, but are  represented instead by single numbers which are meant to summarize an important property of the high-dimensional set, for example, the orientational preference, or the occularity. The models themselves are generally not meant to explain neurophysiological mechanisms, however, certain generalities of neural organization can be discovered from their results.

\subsection{High-Dimensional Models of Receptive Fields and V1 Self-
Organization}

Both the development of orientational selectivity by individual neurons 
and the organization of V1 into feature maps are included in these 
models. Receptive fields are defined in terms of  connection strengths 
between the LGN and V1, and the connection strengths are learned on 
the basis of correlations in the input data set. These are the models of 
von der Malsburg et al\cite{MALS73,MALS77A,MALS81}, Linsker\cite{LINS86,LINS90A,LINS86A}, Miller\cite{MILL89,MACK90,MILL94}, and Obermeyer, et al\cite{OBER91}. The last is called a competitive hebbian model by Erwin, et al\cite{ERWI94}, because the learning is hebbian and the lateral interaction responsible for the feature map formation may be  interpreted as competition between V1 neurons. The former two models, and the present work, are of the type referred to as correlation-based learning models. The high-dimensional models are meant to be more biologically realistic, and succeed in this to varying degrees. Below we review a few of the other high-dimensional models.

\subsubsection{von der Malsburg}

In 1973, von der Malsburg performed simulations on a model for 
orientational feature map development\cite{MALS73}. He employed 
hebbian learning of connections of both excitatory and inhibitory cells, 
with bar-shaped inputs to the visual field. Lateral interactions in the 
cortical layer were included and the weights' growth was controlled 
through normalization. Though his network was small, an orientational 
feature map clearly developed.

\subsubsection{Linsker}

Linsker developed a feedforward model with overlapping receptive 
fields and hebbian learning of connections\cite{LINS86,LINS90A,LINS86A}. He found that after six successive layers of connections developed in this way, the receptive fields were of the center-surround type. In the seventh layer he obtained oriented bar detectors. Random inputs were approximated by computing an expected correlation matrix for activities in the sixth layer (ensemble averaging) and solving a differential equation for the weight values. A saturation limit on the weight values was imposed. Feature map formation was simulated in a low-dimensional model by assigning to each V1 neuron an angle corresponding to orientational preference and computing the effect of lateral interactions by an annealing process. The lateral interaction was excitatory only. The model does produce an orientational feature map, which is compared to experimentally-obtained maps in ref. \cite{ERWI94}. A correlation between cortical coordinates and  orientation value is seen, a phenomenon which is not observed in experiments.

\subsubsection{Miller}

Miller's model\cite{MILL89,MACK90,MILL94} is mathematically similar 
to Linsker's final stage, but with a more physiological interpretation. 
Forward connection strengths are learned, with connections to both ON-
center and OFF-center geniculate cells. A difference-of-gaussians lateral 
interaction is assumed. As in Linsker's model, ensemble averaging is 
employed, and the differential equation for the weights is solved 
numerically. The correlation function for the ensemble average is 
assumed to be of the center-surround type and is explained as the 
result of competition between geniculate cells. Weight magnitudes are 
controlled by subtractive terms in the differential equation, 
renormalization, and a saturation limit. The results are similar to the 
results of Linsker's model\cite{ERWI94}.

\subsubsection{Obermeyer, Ritter, and Schulten}

Both low-dimensional\cite{OBER90A,OBER90C,OBY-91B,OBY-92C} and 
high-dimensional\cite{OBER91} models were studied, based on 
Kohonen's algorithm for self-organizing feature maps\cite{KOHO82B}. 
Kohonen's algorithm may be considered as an approximation to 
employing a difference-of-gaussians lateral interaction. Instead of 
connections, an exponential neighborhood function is used which represents the effect on neural activities that a lateral interaction would have\cite{RITT91A}. The high-dimensional model employs connections only to
ON-center cells. Receptive fields of elongated shape are used to 
represent orientational preference. The input set consists of ovals of 
various positions and ellipticities. Retinotopicity, ocular dominance, and 
orientational preference and specificity were all included in the model. 
The model yields good comparison to experiment on all accounts, 
however, geniculate receptive field sizes and OFF-center cells were not 
taken into account, and it is not clear what the results would be with a 
more realistic input data set.

\section{Realistic Theories}

At this point, we would like to describe what we feel qualifies a model 
as being biologically realistic. In the past, when computers were  much 
slower and smaller than those of today, it was impossible to execute a 
realistic large-scale simulation of a brain area. As such work becomes 
possible, a discussion of the standards for neurophysiological realism is 
in order. 

In order to be biologically realistic, a model must:

1) Be high-dimensional, that is, the basic quantities of the model are connection strengths and neural activities. Quantities such as "orientation", measured as an angle, are not physical quantities, and so must be derived from the output of the model.  In the present work, it will be shown that this approach results in a functioning model of the visual system which can be used to process images. 

2) Use plausible learning only\cite{HEBB49A}. Exotic learning rules that 
cannot have a plausible implementation in brain should be avoided.
Such a constraint lies at the heart of explanatory brain modelling; the 
object is not to simply show that the function can be done by some 
system, but to propose how the brain actually performs the function 
with the physiological capabilities that it has. When the modeler so 
limits his own set of tools, he is forced to discover a method that at least 
has a chance of being found in nature. We employ hebbian learning 
as a plausible learning mechanism.

3) Learn all but the simplest connections.  "Prewiring" of connections 
should be avoided. In other words, the network should be "self-
organizing"; only the raw sensory inputs and the learning rule should be 
enough for the network to acquire its needed properties. Some 
connections in the brain, though, could be formed through other than a 
learning process. If they are made, for example, by a diffusion 
mechanism, this process might not be included in a neural network 
model.

4) Use only neuronal types and connection pathways that are suggested 
by physiology. This is virtually the only degree of freedom left to the 
modeler, but even this must be limited by the current state of 
knowledge. In fact, much of the experimental research effort so far has 
gone into the exploration of these components of neural anatomy. 
Therefore these studies constrain, but also guide, the construction of 
theories. Nevertheless, gaps in the understanding of the brain in this 
regard must be filled in by the modeler, and such 
proposals constitute the contribution of the theory.

\section{Description of the Theory}

We have attempted to fulfill these requirements in our model. The goal 
of our model is to describe the development of the receptive fields of 
V1 neurons ant their organization in the visual cortex. The model 
incorporates many details of the visual system which have been 
discovered experimentally; included in the model are the functions of 
the retina, the lateral geniculate nucleus, and the primary visual cortex. 
The sizes and overlapping character of the receptive fields of neurons in 
the visual system are important physiological properties which are also 
components of the model. While conforming to the most important 
physiological constraints, the model at the same time implements  the 
simplest possible mathematical principles in which the algorithmic 
description of the dynamics of the network can be formulated. The aim 
is that the performance of the network be dependent as much as 
possible on the above-mentioned physiological composition, further 
contributing to the goal of biological realism. In the end, powerful 
principles of information processing emerge, such as a maximizing of 
the rate of information transfer through the visual pathway,  suggesting 
that the natural system on which the model is based is an optimal 
system in regards to its function.

We choose to model the following  subsystem: the X-type cells and their projections from the retina to the lateral geniculate nucleus, as  far as the primary visual cortex (area V1), in the last of which are included  simple cells consisting of one excitatory and one inhibitory region, e.g.,  oriented 
edge detectors.  We take as the function of this processing pathway the 
task of representing a  visual image by the activities of neurons in the 
primary visual cortex, in a way  which is compatible with their 
receptive field properties.  

\subsection{Elements of the Model}

The model contains the following elements:

\begin{enumerate}
\item Retinal photoreceptors and their connections to retinal ganglion 
cells

\item  Center-surround retinal ganglion cells.

\item  Forward excitatory 
connections from the retina to the lateral geniculate nucleus, or LGN.

\item  Center-surround cells of the lateral geniculate nucleus.

\item  Forward excitatory connections from the LGN to the primary 
visual cortex, or V1.

\item  Simple cells having orientational preference in the primary 
visual cortex. 

\item  Lateral connections between cells in V1.

\item  Backward inhibitory connections from V1 to the LGN.

\end{enumerate}

The model's mechanism for the development of  V1 is described by the 
learning of the forward and backward connection strengths between the 
LGN and V1. We continue by defining the interaction of these elements 
and the learning of connection strengths via mathematical equations.

\subsection{Neural Activities and Learning Rules}

Let $x_i$ represent the input to a photoreceptor at position $i$, and $y_j$, and $a_k$ the activities of a retinal ganglion cell at position $j$, and a V1 neuron at position $k$, in their respective neural layers. Since projections from layer to layer are retinotopic,  the subscripts also indicate the position 
of the centers of their receptive fields in visual space. Activities are 
defined in terms of average firing rates. Negative activities of a neuron are allowed and are interpreted as values below a level of "idling" activity. They could also  be  interpreted as activities of another neuron of opposite response characteristics at the same position.

Input images falling on the photoreceptors of the eye are filtered by 
center-surround retinal ganglion cells. We put
\begin{equation}\label{eq:lgnact}
y_j = \sum_i  g_{ij} x_i
\end{equation}
where $g_{ij}$ represents the center-surround receptive field of the retinal ganglion cell  $j$. $g_{ij}$ depends only on the distance of $i$ from $j$ in the retina, and is modeled by a difference-of-gaussians function of average value zero. The function is the same for every $j$. This convolution removes most of the broad intensity information from the image. Information about edges remains. The retinal ganglion cells are connected to LGN relay neurons in a one-to-one manner and excite the LGN relay neurons.  Therefore we take $y_j$ to also be the initial activity of the LGN relay neuron whose receptive field is at position $j$ in the visual field. There are feedforward connections from the LGN to V1, and lateral connections between V1 neurons. These connections are represented by the values $w_{jk}$ for the feedforward connection between LGN relay neuron $j$ and V1 neuron $k$, and $h_{kl}$  for the lateral connection between V1 neurons $k$ and $l$. $h_{kl}$ is of similar form to $g_{ij}$; the lateral interaction is excitatory to nearby neighbors and inhibitory to neighbors farther away. The activity of neurons in layer V1 is given by:
\begin{equation}\label{eq:v1act}
a_l = \sum_{j,k} h_{kl} w_{jk} y_j
\end{equation}
i.e., the forward and lateral interactions are applied in succession.

Reciprocal connections also exist from V1 to the LGN, and the LGN 
activities are inhibited through these backward connections. The 
backward connection between V1 neuron $k$ and LGN relay neuron $j$ 
has the same strength as the forward connection. This is not an 
assumption of the theory, but a conclusion, and later we will give 
several reasons as to why this is reasonable result. The inhibition 
decreases the activity of the LGN relay neuron:
\begin{equation}
\label{eq:inhib}\dot{y}_j = -\sum_k  w_{jk} a_k
\end{equation}
As will be shown later, the learning rule is such as to maximize the rate of 
information transfer from the LGN to V1 while the inhibition minimizes the 
activities of the LGN and V1.

The connection strengths $h_{kl}$ and $g_{ij}$ are constant and the
difference-of-gaussians function defining them is an assumption of the 
model. The connections $w_{jk}$, both forward and backward, are 
learned by a hebbian rule relating the activities of the LGN and V1 
neurons:
\begin{equation}\label{eq:hebbrule}
\dot{w}_{jk} =  \eta y_j a_k
\end{equation}
$\eta$ is a decreasing function of time. The receptive fields of the V1 
neurons are determined by their connections $w_{jk}$ to the LGN. We 
can show through simulations that orientation-specific receptive fields
are acquired by the V1 neurons when the hebbian learning is 
implemented. 

The main hypothesis of our work centers on the backward connections 
from V1 to the LGN. Our theory is that the connections result from 
hebbian learning (eq. \ref{eq:hebbrule}) and are connected to inhibitory 
interneurons in the LGN where they terminate. It is known 
experimentally  that these connections are as numerous as the forward 
connections, are retinotopic, and that they terminate
on the inhibitory interneurons in the LGN layer\cite{ORBA84}. Since 
they both learn from the same hebbian inputs, the forward and 
backward connections take on the same form. Later, we propose a 
neural circuit which can account for this formation.

\subsection{The Computational Method}

All the neural layers are modelled by square arrays of cells. We model 
the foveal region by an array of photoreceptors of dimension 304 
by 304. In all layers, a movement  from a cell to a nearest neighbor(one 
neural spacing unit) corresponds to a movement of one pixel in the 
input image, or 0.01 degrees in the visual field. Retinotopy is preserved 
throughout the model by having the index of a neuron in its array also 
represent the position of the neuron's receptive field in visual  space. 
Thus the model subtends 3.04 degrees of arc in the visual field.

\begin{figure}[p]
\vspace{-2.0cm}
\begin{center}
\begin{tabular}[t]{c}
\subfigure[]{\parbox{0.45\textwidth}{
\epsfxsize=0.45\textwidth
\epsfysize=0.45\textwidth
\epsffile{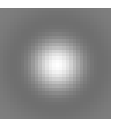}
}
}
\subfigure[]{\parbox{0.45\textwidth}{
\epsfxsize=0.45\textwidth
\epsfysize=0.45\textwidth
\epsffile{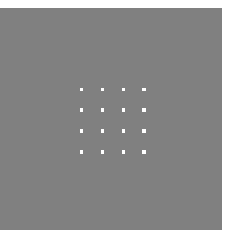}
}
}
\vspace{1.0cm}\\
\subfigure[]{\parbox{0.45\textwidth}{
\epsfxsize=0.45\textwidth
\epsfysize=0.45\textwidth
\epsffile{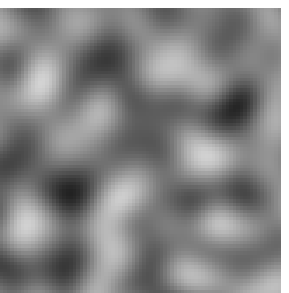}
}
}
\subfigure[]{\parbox{0.45\textwidth}{
\epsfxsize=0.45\textwidth
\epsfysize=0.45\textwidth
\epsffile{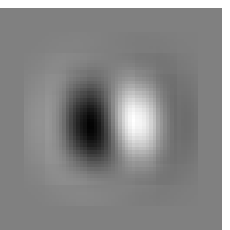}
}
}
\end{tabular}
\end{center}
\vspace{-0.5cm}
\caption[]{  a) Difference-of-gaussians function used to model center-surround receptive field properties of retinal and geniculate cells. The actual size of the array is 32 by 32. b) Grid of 16 LGN neurons connected to each V1 neuron. Position of grid in LGN corresponds to position of V1 neuron in V1. Grid spacing is 6 units for feature maps analysed in this work. c) Typical (equivalent) 320 by 320 image used as input to the LGN. Image is obtained by convolving white noise with center-surround function. d) Typical receptive field of V1 neuron produced by the algorithm. Actual array size is 64 by 64.}
\label{fig:p:mask}
\end{figure}

The receptive field of each retinal ganglion cell has a center radius of 8 
spacing units in the difference-of-gaussians function, and a surround 
radius of 10 units.  See figure \ref{fig:p:mask} a. The receptive field fits into a 32 by 32 array. We do not wish to include retinal ganglion cells whose receptive fields lie only partially within the 304 by 304 array of  photoreceptors. Therefore the size of the array of retinal ganglion cells is 272 by 272.

The connections between the retinal ganglion cells and the cells of the 
lateral geniculate nucleus are assumed to be one-to-one and 
excitatory. Therefore the lateral geniculate nucleus is also a 272 by 272 
array and we model the initial activities of the LGN cells as a direct copy 
of the activities of the retinal ganglion cells. The receptive field of a LGN 
neuron is the same as the retinal ganglion cell corresponding to it.

Each V1 neuron is connected to sixteen LGN neurons, with both 
feedforward and feedback connections to the same LGN cells. The 
sixteen LGN neurons are arranged in a square grid in the LGN layer with 
a spacing of six neural spacing units between grid points, fig. 
\ref{fig:p:mask} b. The size of the grid is chosen so that the receptive 
field diameter approximates the actual value (0.25 
degrees\cite{HUBE74B}) for the foveal region. The connection strengths 
between the LGN and V1 are initialized to random values.

For each V1 neuron, the receptive fields of the sixteen LGN neurons 
taken together fit into a 48 by 48 pixel window in the visual field. In 
order that the receptive field of each V1 neuron lie completely within 
the 304 by 304 input image, the V1 array is of dimension 256 by 256. 
Neuron (0,0) of V1 has its receptive field centered at position (24,24) in 
the visual field. 

The development of the orientational feature map occurs primarily prenatally in 
some species such as the macaque  monkey\cite{WIES74A} and 
postnataly in others. In the latter case, digitized images of natural 
scenes are the inputs to the photoreceptors of the model. For prenatal 
development, we take as the inputs random noise of flat distribution in 
the retina, occurring as spontaneous activity of retinal photoreceptors.  This causes activity in the retinal ganglion cells which is translated to activities of LGN neurons. These activities are calculated according to eq. 
\ref{eq:lgnact}. An example of the resulting LGN activities for a white-
noise photoreceptor output  is shown in fig \ref{fig:p:mask} c .
V1 cells are excited via the feedforward connections and their activities 
are modified by the lateral interactions. The activities of the V1 cells are calculated according to eq. \ref{eq:v1act}. The lateral connections are modelled by a difference-of-gaussians function, of center radius 8 neural spacing units and surround radius 11.8. They extend 32 neural spacing units in each direction from every V1 neuron, except at the edges of the array, 
where the lateral interaction stops. The lateral interactions are iterated 
only once per input image.

The backward connections serve to inhibit the activities of the LGN 
neurons. After the input image is presented and the V1 activities 
calculated, each LGN neuron's activity is decreased by an amount 
proportional to the quantity in eq. \ref{eq:inhib}.

The weights are then updated by Hebb's rule in an amount proportional 
to the quantity in eq. \ref{eq:hebbrule}. Then the above-described process is repeated with a new image. The  backward connections are assumed to also develop by the hebbian rule, but with an inhibitory synapse. Therefore, since they learn from the same inputs, the forward and backward weights have the same values, but opposite sign.  The weight updates occur at different places in the brain; we assume that the neural activities change slowly enough in relation to the learning rate that propagation delays need not be included.

\section{Results of the Simulations}

\subsection{Orientationally Selective Receptive Fields}

Application of this algorithm over many input images will cause the 
weights of each V1 neuron to converge to a configuration exemplified in 
fig. \ref{fig:p:mask} d. The shape represents the first  principal component 
of the set of inputs. The first principal component of a data set has the 
property that the inner product of it with elements of the data set has a 
greater variance than any other component. Since the neuron sums its  
inputs linearly, this just means that the output of the neuron in V1 has a 
greater variance over the input data set than with any other set of 
connections. Therefore the neuron's output encodes the maximal amount 
of information about the activities in  the LGN. For if one considers some 
finite resolution for sensing the activity of the V1 neuron, then the 
larger the range of the activity, the more the possible distinguishable 
values the activity can take, and therefore, according to Shannon's 
measure of information, the more information can be encoded in each 
activity value.

\subsection{Dependence of Receptive Field Type on LGN Spacing}

\begin{figure}[p]
\begin{center}
\begin{tabular}[t]{c}
\subfigure[]{\parbox{0.45\textwidth}{
\epsfxsize=0.45\textwidth
\epsfysize=0.45\textwidth
\epsffile{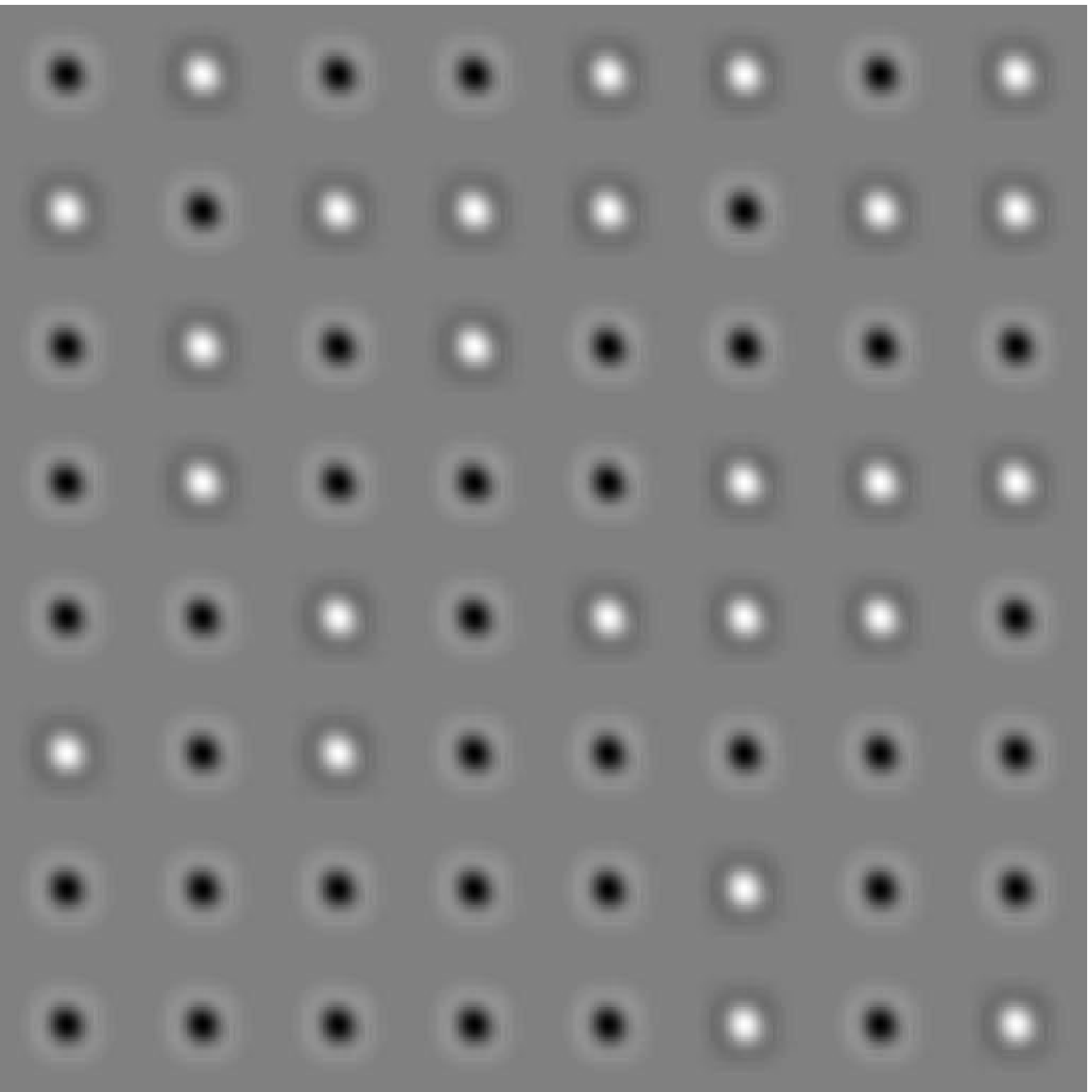}
}
}
\subfigure[]{\parbox{0.45\textwidth}{
\epsfxsize=0.45\textwidth
\epsfysize=0.45\textwidth
\epsffile{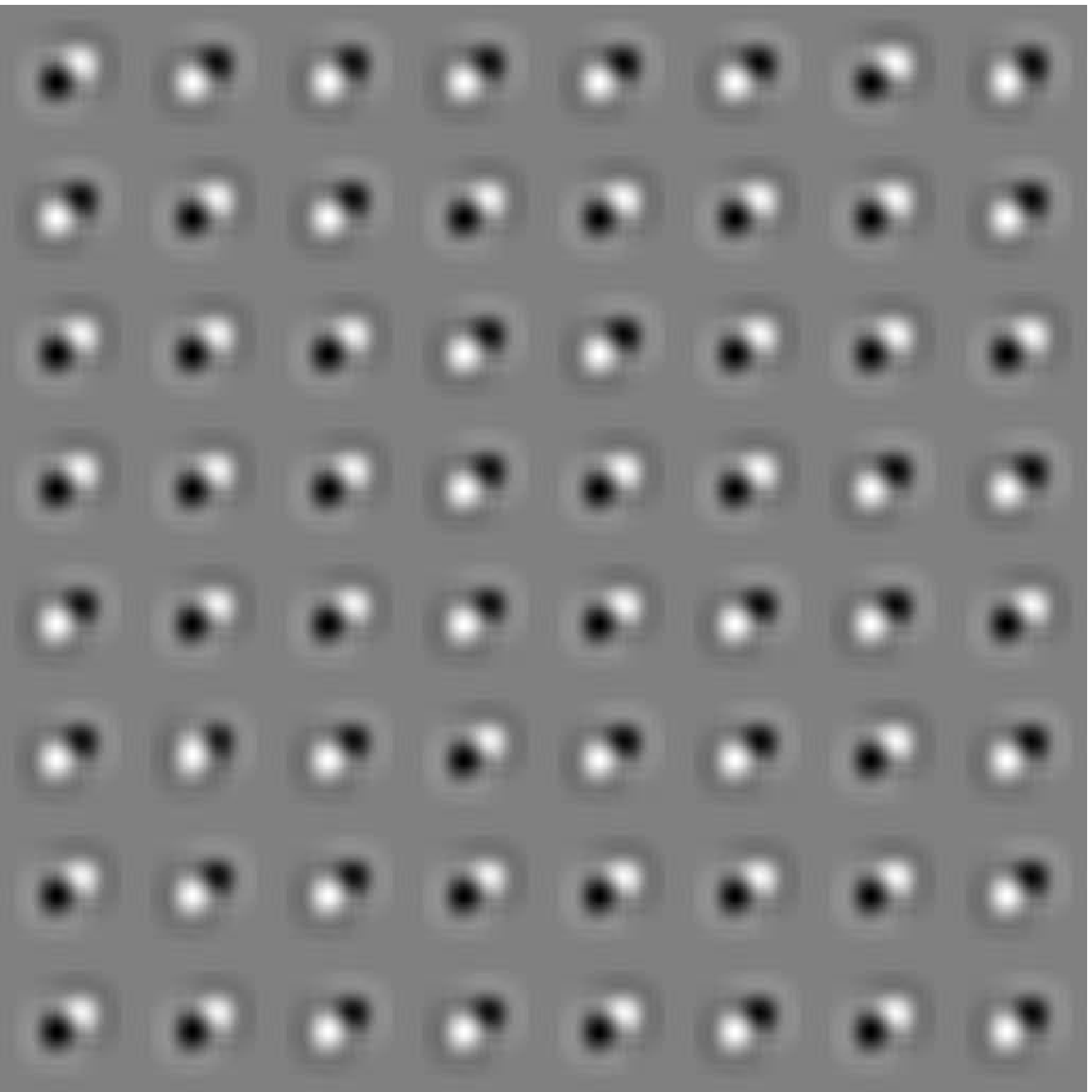}
}
}
\vspace{1.0cm}\\
\subfigure[]{\parbox{0.45\textwidth}{
\epsfxsize=0.45\textwidth
\epsfysize=0.45\textwidth
\epsffile{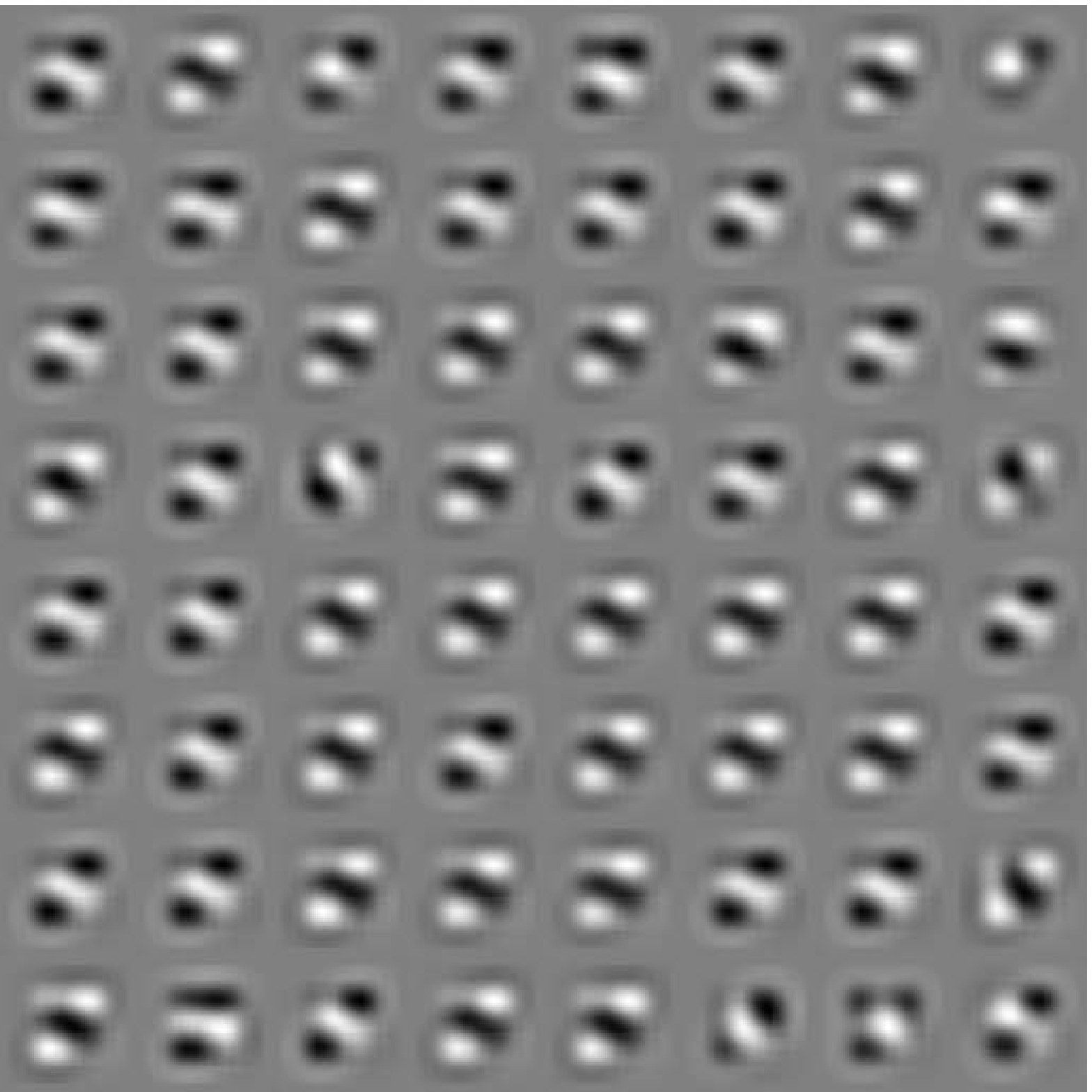}
}
}
\subfigure[]{\parbox{0.45\textwidth}{
\epsfxsize=0.45\textwidth
\epsfysize=0.45\textwidth
\epsffile{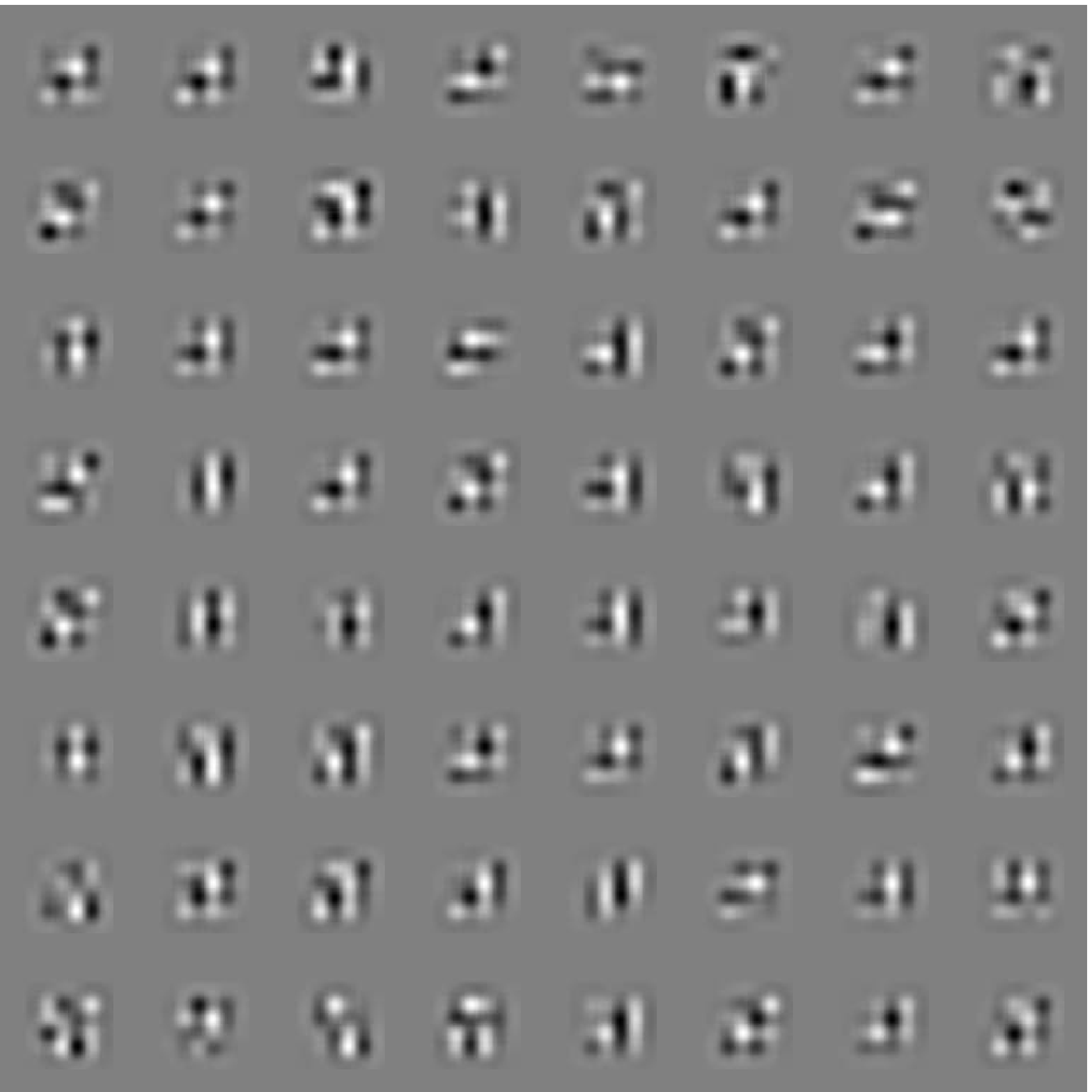}
}
}
\end{tabular}
\end{center}
\vspace{-0.5cm}
\caption[]{{8 by 8 array of receptive fields developed for various arborization densities. All neurons have 16 connections to LGN cells in a 4 by 4 grid with spacing a) 4 units b) 6 units c) 8 units d) 10 units}}
\label{fig:p:rfs}
\end{figure}

The choice of spacing between the centers of LGN receptive fields connected to a V1 neuron determines the type of receptive field that is formed. If the spacing is small in comparison with the correlation length of LGN activities, the receptive field takes on a circularly symmetric shape\ref{fig:p:rfs} a. The receptive field has a center-surround shape due to the surround portions of the LGN receptive fields that comprise it. At greater spacings, the receptive field has more subfields, becoming a detector of edges, bars, and finally spatial frequencies as the spacing increases. Thus the statistical mix of densities of LGN cells afferent to individual V1 neurons determines the statistical mix of receptive field types found in V1.

\subsection{Orientational Feature Map}

\begin{figure}[p]
\vspace{-2.0cm}
\begin{center}
\begin{tabular}[t]{c}
\subfigure[]{\parbox{0.45\textwidth}{
\epsfxsize=0.45\textwidth
\epsfysize=0.45\textwidth
\epsffile{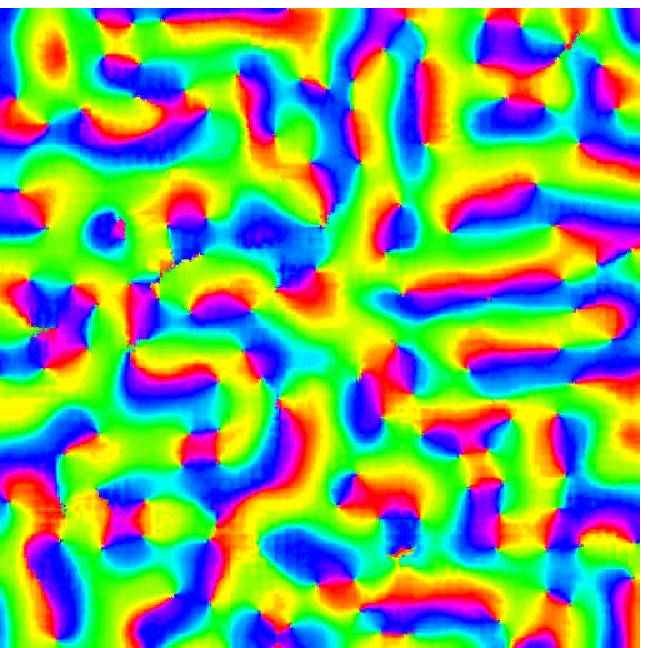}
}
}
\subfigure[]{\parbox{0.45\textwidth}{
\epsfxsize=0.45\textwidth
\epsfysize=0.45\textwidth
\epsffile{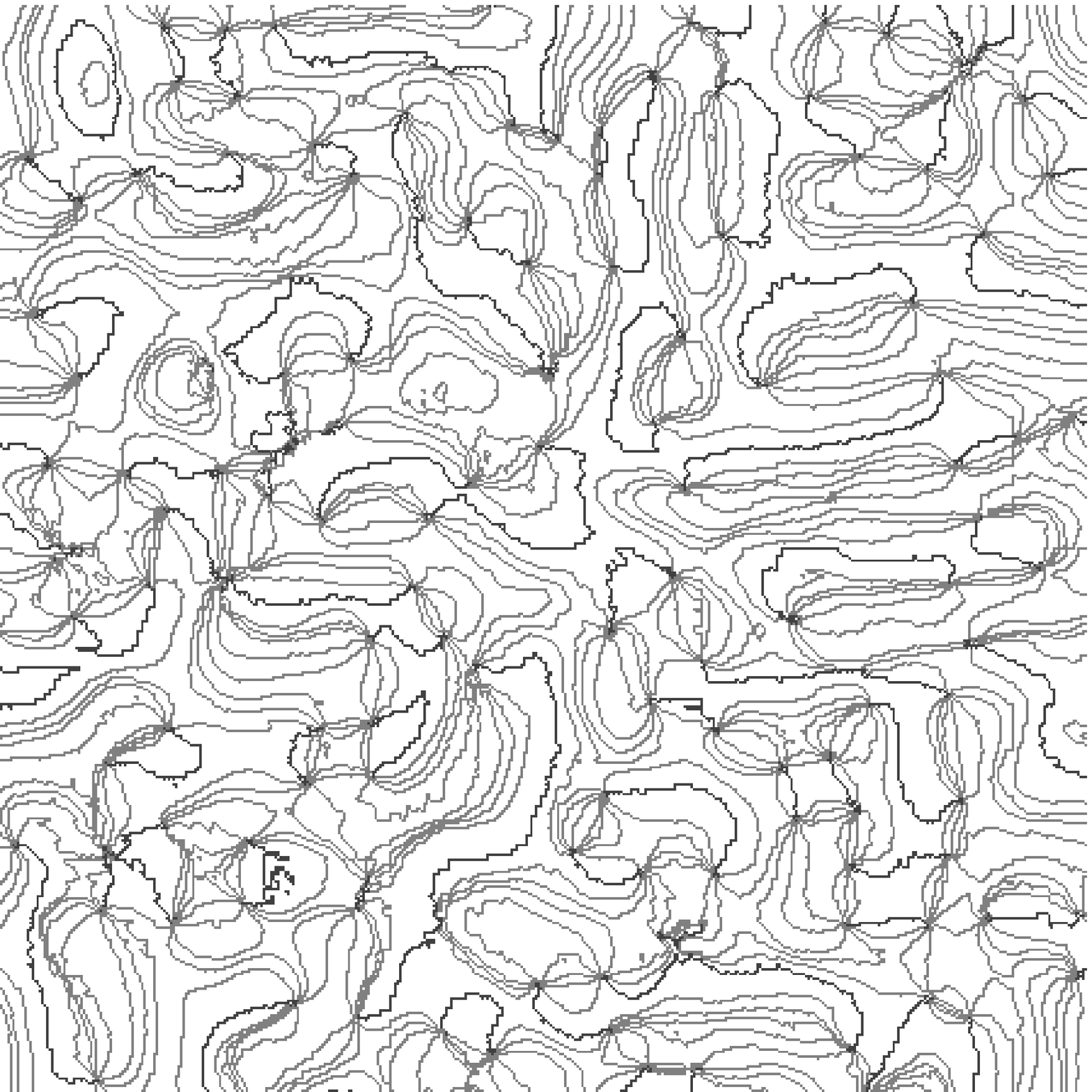}
}
}
\vspace{1.0cm}\\
\subfigure[]{\parbox{0.45\textwidth}{
\epsfxsize=0.45\textwidth
\epsfysize=0.45\textwidth
\epsffile{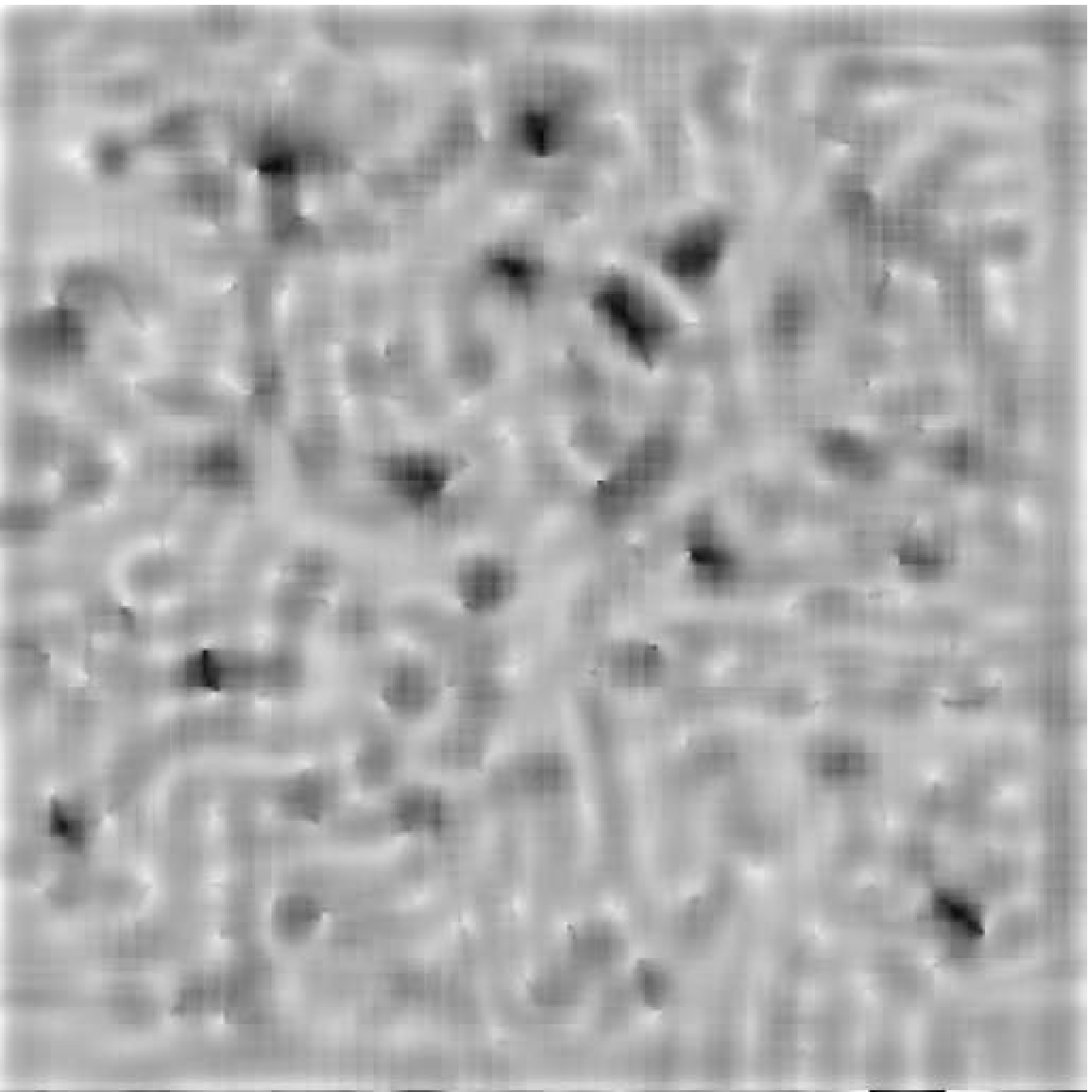}
}
}
\subfigure[]{\parbox{0.45\textwidth}{
\epsfxsize=0.45\textwidth
\epsfysize=0.45\textwidth
\epsffile{figures/5120/grad.eps}
}
}
\end{tabular}
\end{center}
\vspace{-0.5cm}
\caption[]{  a) Orientation column arrangement after 5120 steps. The section size is 224 $\times$ 224.  Each pixel in the illustration corresponds to one orientation column in the primary visual cortex. The orientational preference of the column is represented by the color of the pixel. b) Iso-orientation contours of the orientational feature map. Line spacing is 20 degrees of orientational preference. c) Specificity of orientational preference. The specificity is the response of the V1 neuron to an edge of optimal orientation. d) Gradient of orientational preference. Dark narrow lines indicate fractures.}

\label{fig:p:map}
\end{figure}

\begin{figure}[p]
\vspace{-2.0cm}
\begin{center}
\begin{tabular}[t]{c}
\subfigure[]{\parbox{0.45\textwidth}{
\epsfxsize=0.45\textwidth
\epsfysize=0.45\textwidth
\epsffile{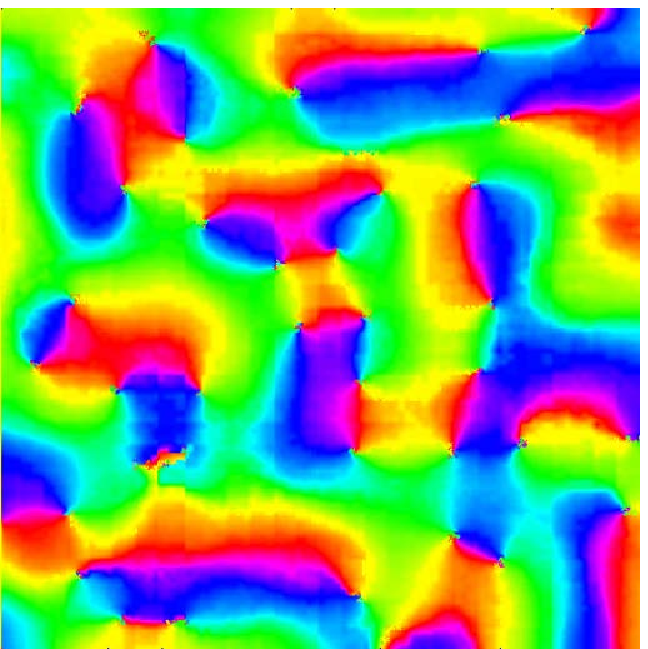}}
}
\subfigure[]{\parbox{0.45\textwidth}{
\epsfxsize=0.45\textwidth
\epsfysize=0.45\textwidth
\epsffile{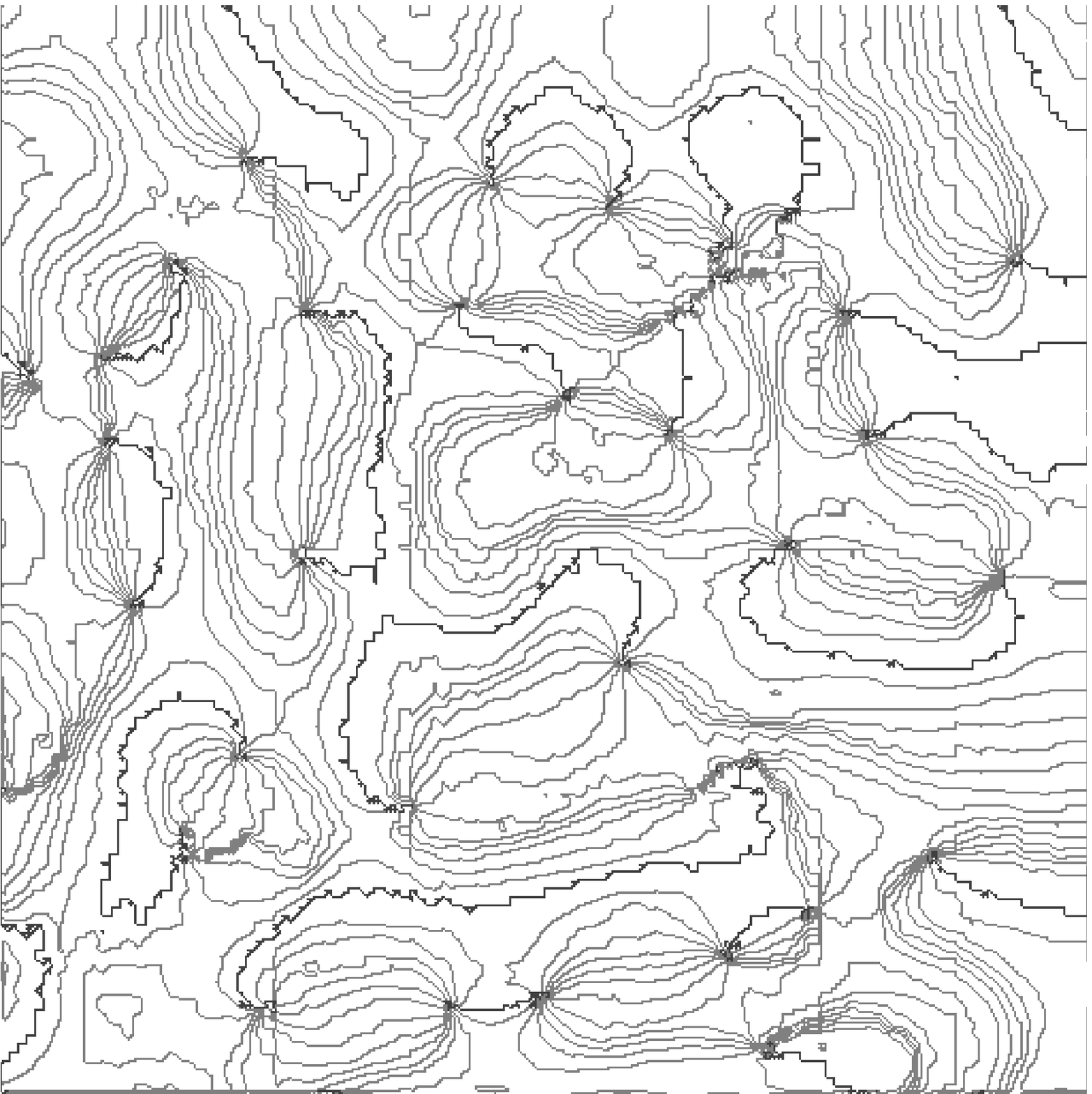}
}
}
\vspace{1.0cm}\\
\subfigure[]{\parbox{0.45\textwidth}{
\epsfxsize=0.413\textwidth
\epsfysize=0.45\textwidth
\epsffile{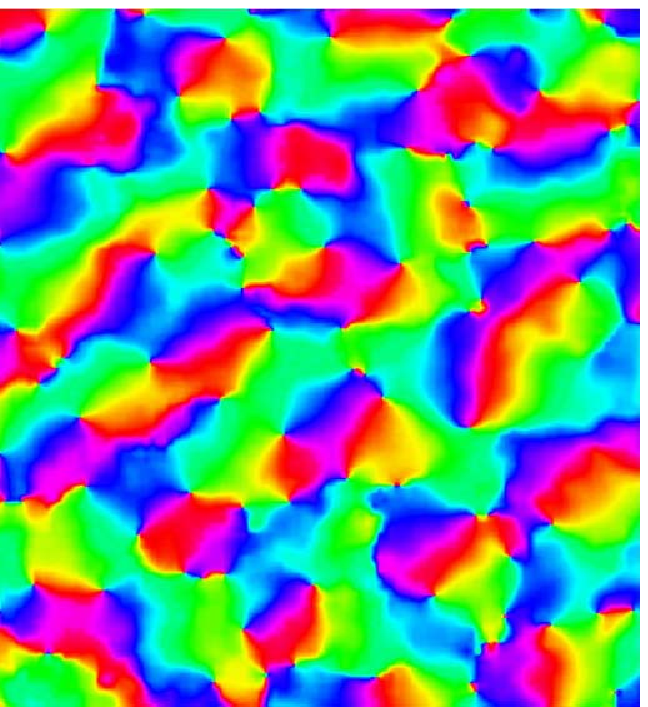}
}
}
\subfigure[]{\parbox{0.45\textwidth}{
\epsfxsize=0.413\textwidth
\epsfysize=0.45\textwidth
\epsffile{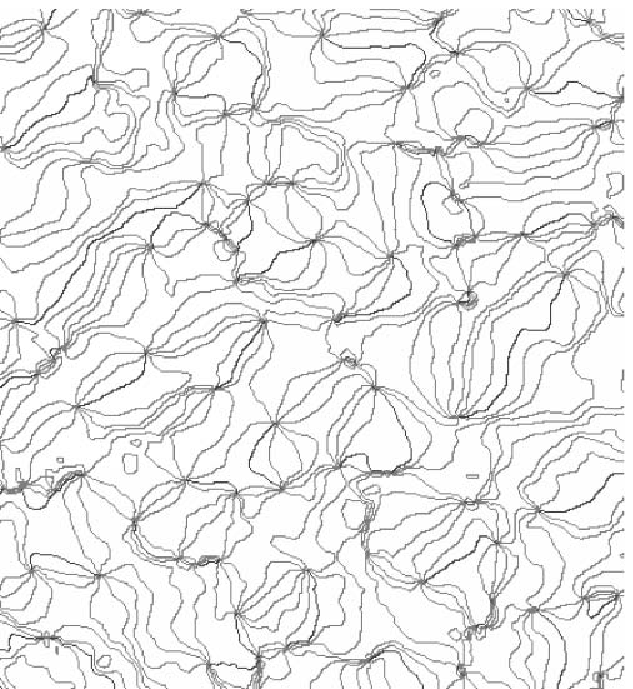}
}
}
\end{tabular}
\end{center}
\vspace{-0.7cm}
\caption[]{  a) Orientation column arrangement after 5120 steps. Lower right-hand corner of full orientational feature map.  Each pixel in the illustration corresponds to one orientation column in the primary visual cortex. The orientational preference of the column is represented by the color of the pixel. b) Iso-orientation contours of the feature map. c) Orientation column arrangement from experimental data of macaque monkey. The array size is 448 $\times$ 480. d) Iso-orientation contours of experimental data.}

\label{fig:p:mapsmall}
\end{figure}

Most V1 neurons' receptive fields are of the form shown in fig. 
\ref{fig:p:mask} d, showing that the response of the neuron would show a preference for stimuli of a preferred orientation. The orientation of the receptive field depends on the lateral interactions included in the model. The orientational feature map is shown in figure \ref{fig:p:map} a, and an iso-orientation plot of the same map is shown in figure \ref{fig:p:map} b. An enlargement of the lower-right quadrant of the map is shown in figure 
\ref{fig:p:mapsmall} a, and can be compared with a feature map 
obtained from experiment in figure \ref{fig:p:mapsmall} c. 

Erwin, et al\cite{ERWI94},analysed many feature map models and arrived at a 
series of tests to compare simulation results to experiments for both 
orientation and ocular dominance. Only the former property is 
considered in the present work. For orientational column structure, the 
following properties are considered and the present model is compared 
to experiment.

\subsubsection{Singular Points, Linear Zones, and Saddle Points}

Orientational feature maps observed in monkey experiments show 
stripes of common orientational preference. Stripes of all orientations 
converge at numerous singular points scattered through the feature 
map. Conversely, singular points are commonly connected by stripes of uniform orientational preference. A region of the cortex over which all orientational preferences occur is called a hypercolumn. Linear zones are defined as regions in which traversing along a straight line parallel to the cortical sheet yields a continuous linear change in orientational preference. 
These regions exist wherever stripes of orientational preference run
parallel, or nearly parallel, to each other. These regions occur between 
most neighboring pairs of singular points, when the singular points are 
connected by orientational preference stripes covering a certain range 
of orientations. These structures can be seen in the simulation 
results. Saddle points occur between linear zones and are as defined in 
analytic geometry on the surface obtained by plotting the orientational 
preference as a function of cortical position. These can be seen most 
easily in the colored map of the simulation results fig. \ref{fig:p:map}a 
as square-shaped areas of color.

\subsubsection{Specificities}

Specificity is the strength of a neuron's response to an optimally-
oriented input. Experiments have found that the specificity tends to be 
less near the singular points of the map. We found this to be true for 
many singular points of our map, especially early in the simulation. At 
early times, the low specificity at singular points is caused by conflicting 
influence of neighboring neurons due to the variety of orientations 
found there. Receptive fields near the singular point are "pulled" 
towards one orientation and then another, different, one when the input 
changes. The effects cancel, and the receptive field tends to resemble a 
shape which is independent of all the oriented receptive fields. In our 
simulations, the receptive fields even near singular points do eventually 
become stable at the typical shape. (fig. \ref{fig:p:mask} d). The map's 
specificity values are shown in fig. \ref{fig:p:map} c.

\subsubsection{Fractures}

\begin{figure}[p]
\vspace{-2.0cm}
\begin{center}
\begin{tabular}[t]{c}
\subfigure[]{\parbox{0.45\textwidth}{
\epsfxsize=0.45\textwidth
\epsfysize=0.45\textwidth
\epsffile{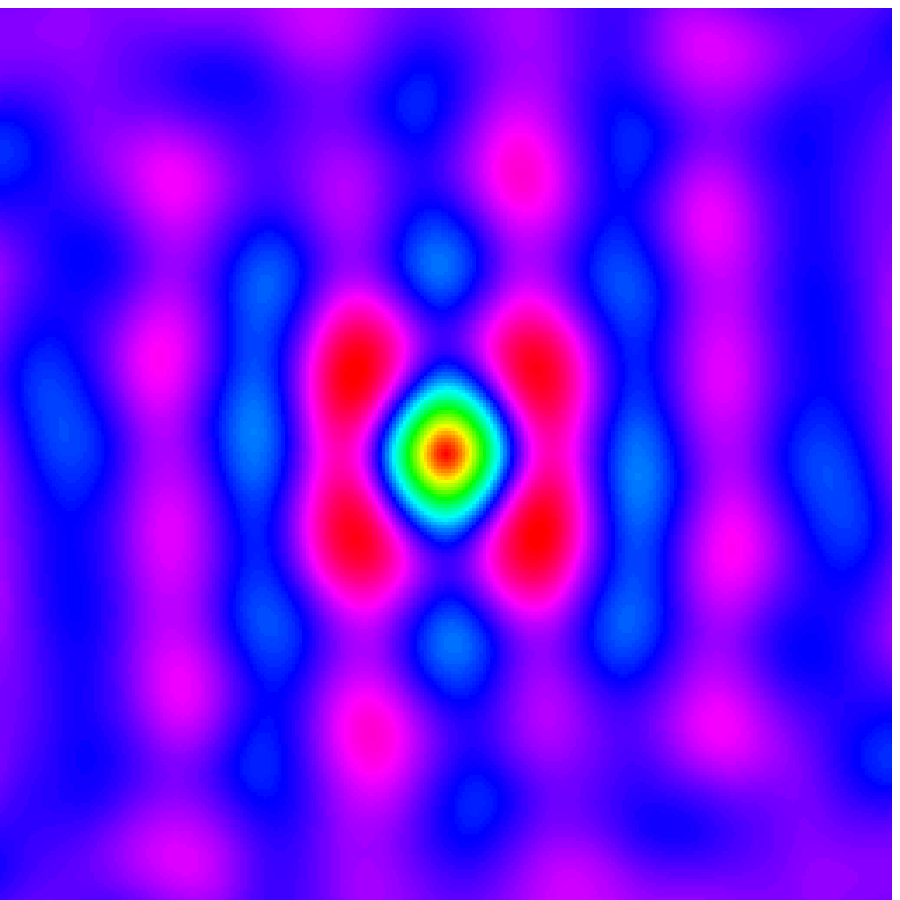}
}
}
\subfigure[]{\parbox{0.45\textwidth}{
\epsfxsize=0.45\textwidth
\epsfysize=0.45\textwidth
\epsffile{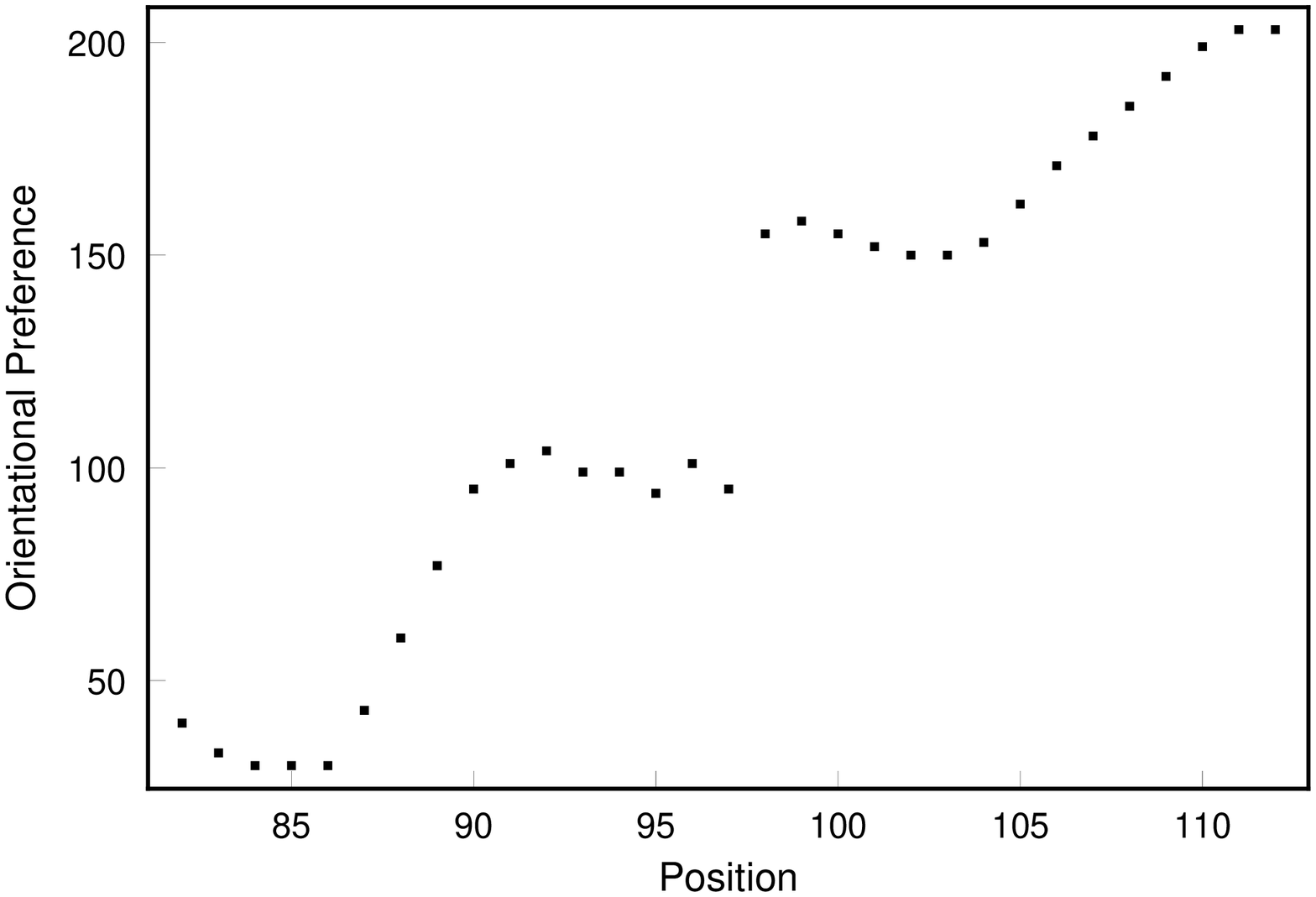}
}
}
\vspace{1.0cm}\\
\subfigure[]{\parbox{0.45\textwidth}{
\epsfxsize=0.45\textwidth
\epsfysize=0.45\textwidth
\epsffile{figures/5120/orhist.ai}
}
}
\subfigure[]{\parbox{0.45\textwidth}{
\epsfxsize=0.45\textwidth
\epsfysize=0.45\textwidth
\epsffile{figures/5120/crhst.ai}
}
}
\end{tabular}
\end{center}
\vspace{-0.5cm}
\caption[]{  a) Two-dimensional correlation function of orientational feature map. b) Simulation of electrode penentration measurement showing fracture in feature map. c) Histogram of orientational preferences. Imbalance toward preferences of horizontal and vertical directions is clearly seen. This is probably due to rectangular grid of LGN neurons used to form V1 receptive fields. d) Histogram of intersection angle between orienational preference and direction of gradient of the orientation. Histogram shows that orientational preference tends to be aligned with the gradient of the orientational feature map.}

\label{fig:p:corr}
\end{figure}

Fractures are lines of discontinuity in the feature maps\cite{BLAS86}. 
They can be seen as the sharp dark lines in the plot of the gradient of the 
orientation distribution, fig. \ref{fig:p:map} c. A one dimensional plot of 
orientational  preference vs position on a line through a fracture is 
shown in  fig. \ref{fig:p:corr} b, and shows both linear zones and the 
fracture.

\subsubsection{Fourier Analysis}

\begin{figure}[p]
\vspace{-1.5cm}
\begin{center}
\begin{tabular}[t]{c}
\subfigure[]{\parbox{0.45\textwidth}{
\epsfxsize=0.45\textwidth
\epsfysize=0.45\textwidth
\epsffile{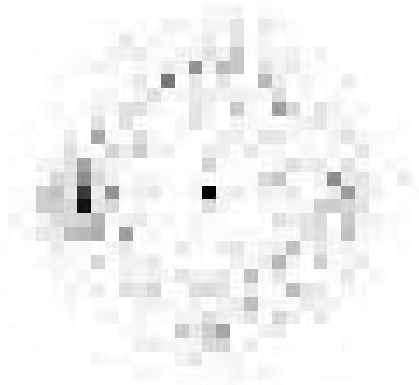}
}
}
\subfigure[]{\parbox{0.45\textwidth}{
\epsfxsize=0.45\textwidth
\epsfysize=0.45\textwidth
\epsffile{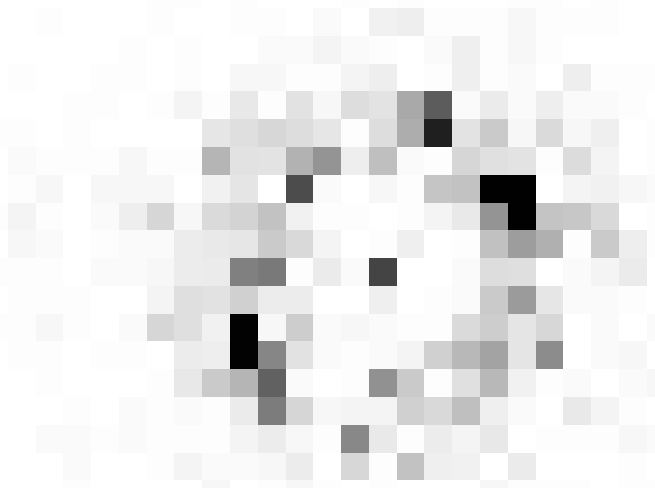}
}
}
\vspace{1.0cm}\\
\subfigure[]{\parbox{0.45\textwidth}{
\epsfxsize=0.45\textwidth
\epsfysize=0.45\textwidth
\epsffile{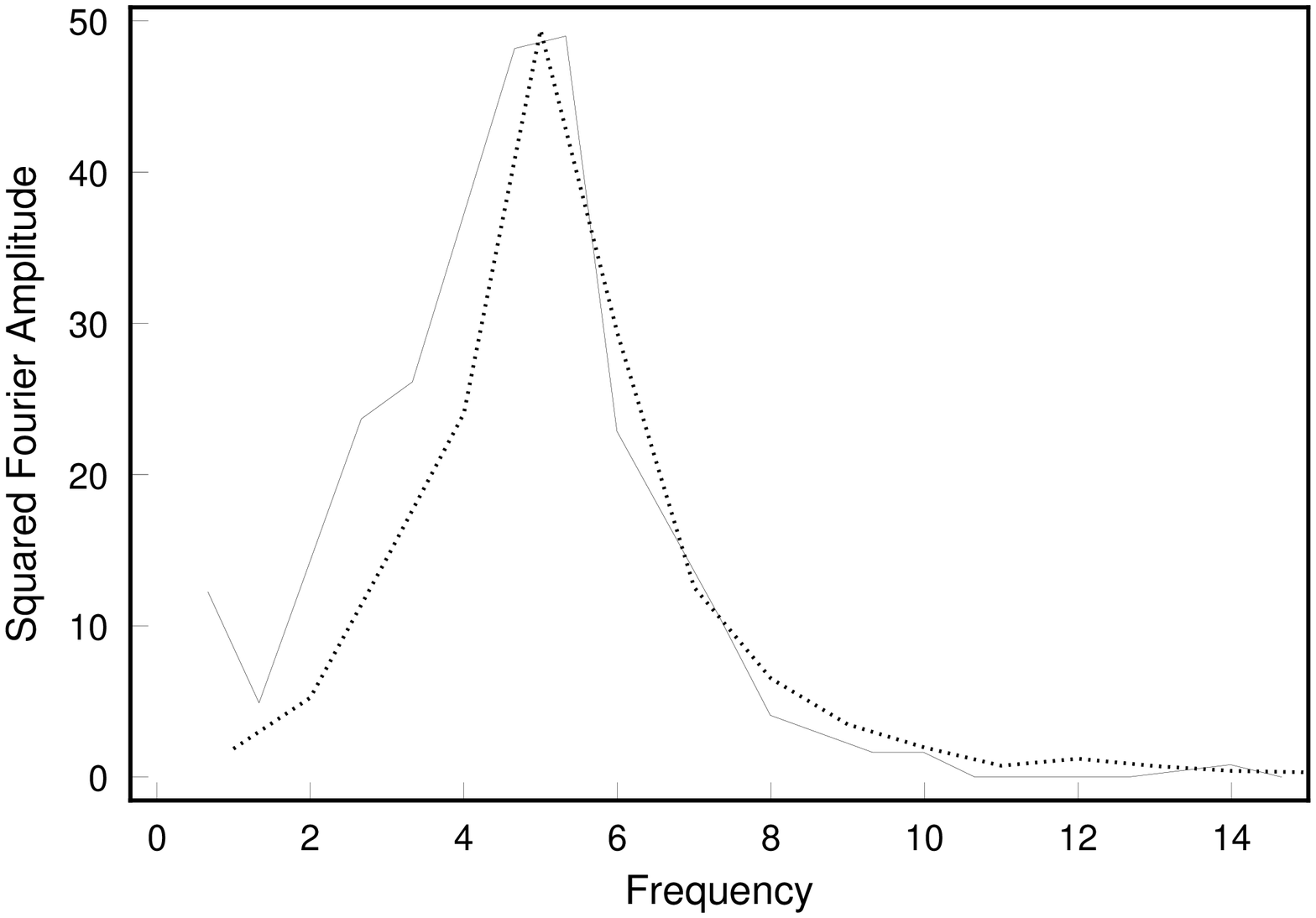}
}
}
\subfigure[]{\parbox{0.45\textwidth}{
\epsfxsize=0.45\textwidth
\epsfysize=0.45\textwidth
\epsffile{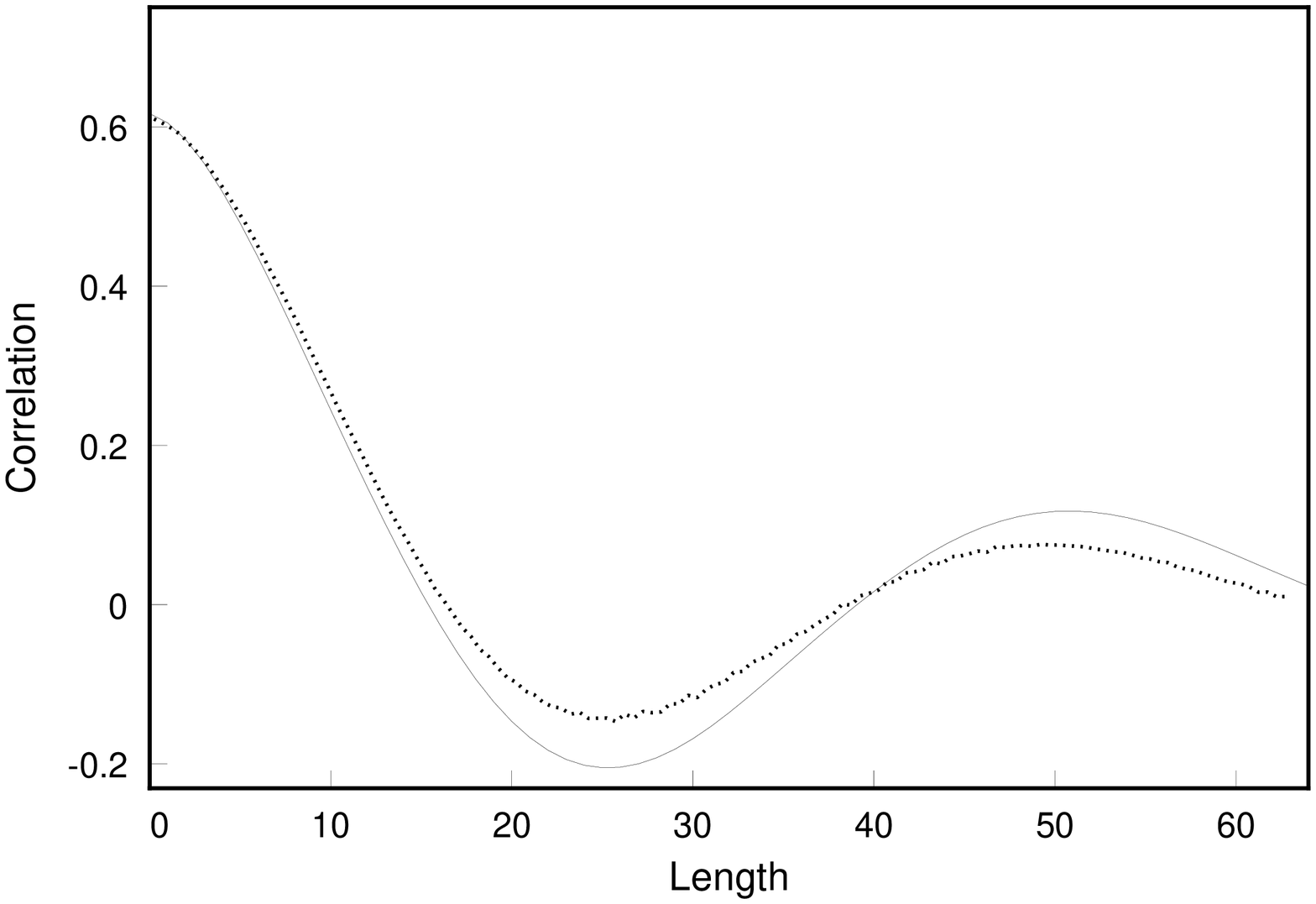}
}
}
\end{tabular}
\end{center}
\vspace{-0.5cm}
\caption[]{  a) Squared amplitudes of fourier transform of orientational feature map. b) Squared amplitudes of fourier transform of map obtained from experiment. c) Average values along along the diagonals of figure a) compared to values from b) averaged over all angles. d) Values along along a half-diagonal of the correlation function of orientational feature map compared to correlation values from experimenatal results averaged over all angles.}

\label{fig:p:four}
\end{figure}

A two-dimensional fourier transform of the feature map shows a 
bandpass-type shape, nearly circularly symmetric. The squared 
amplitudes of the fourier components are shown in fig. 
\ref{fig:p:four}a. There is some strengthening of components in the 
horizontal and vertical directions as an artefact of the grid geometry. 
The squared amplitudes of the fourier components of the 
experimentally-obtained map are shown in figure \ref{fig:p:four}b. The 
energy spectrum taken from an average along the diagonals is 
compared to a radial average from the fourier spectrum of the feature  
map obtained from experiment in fig. \ref{fig:p:four}c. The vertical and 
horizontal axes have been scaled to align the peaks of the two plots. The 
high-frequency tail falls off according to a power law, which is expected 
in the transform of a function which contains singularities . The result 
from the simulation is close to the experimental result here, indicating 
that the two data sets have similar structure near the singularities.

At lower frequencies, the plots show that the bandwidth of the 
simulation results is greater than that of the experimental results. The
orientational stripes from the simulation show greater variation in their 
width than those found experimentally, as can also be seen by 
comparing the iso-orientation contour plots figs. \ref{fig:p:mapsmall}b, 
\ref{fig:p:mapsmall}d. This is consistent with the histogram of orientation distributions, fig.\ref{fig:p:corr}c, which shows more orientational preferences near the horizontal and vertical directions. This is probably a result of the rectangular grid used for the location of LGN neurons afferent to each V1 neuron. The grid accommodates horizontally and vertically symmetric V1 receptive fields more readily than those with symmetries in other directions.

\subsubsection{Correlation Functions}

The orientational feature map can be considered as an array of hypercolumns. Repetition of these units occurs over the long range, but not in an ordered manner such as repeated, duplicate units. Therefore the correlation function of the feature map should fall off with distance.The correlation function of  the orientational feature map does show this behavior. Both the experimental and the simulation results are shown in figure \ref{fig:p:four}d. The two-dimensional correlation function is shown in figure \ref{fig:p:corr}a. The correlation function is taken as a half-diagonal of the two-dimensional function rather than integrating about the angular coordinate because the 2-dimensional correlation function is not circularly symmetric.

The correlation functions have alternating positive and negative regions, 
each of  about half the typical distance between singular  points. The 
positive regions occur at distances where similar orientations can be 
found, and the negative regions arise from distances where orientations 
differing by ninety degrees are expected. The range of orientations is 
one hundred eighty degrees, the sign of the
receptive field being ignored. For purposes of comparison, the data are 
scaled in the vertical axis so that the intercepts of the two plots are 
equal, and in the horizontal axis so that the lengths of the alternating 
regions are the same for both plots.

The figure shows that there is more medium-range order in the feature 
map from the model. This is due to horizontal and vertical correlations 
produced as artefacts of the use of a rectangular grid to represent the 
arrangement of efferent connections from the LGN and to represent the 
cortex itself. There was a tendency for neurons along a horizontal or 
vertical line to lock onto the same orientation preference. This artifact is 
related to the LGN grid described above and to the use of a rectangular 
array to represent the neural layer and, consequently, the lateral 
connections. Specifically, the lateral connections in the horizontal or 
vertical direction are more dense that those in other directions. The 
effect is as high as a factor of 1.414 between the diagonal and the 
vertical or horizontal. The macroscopic patterns thus developed tend to 
have relatively strong horizontal and vertical components. In 
addition, the stripes of the animal brains are more wavy than those of 
the simulation, possibly due to
influences from the presence of blood vessels or deformations of the 
cortical sheet. 

  
 
morphological 
 

   \subsubsection{Independence of Cortical Coordinates and Orientation 
Preference}

   
 
 
 
 
 
 

We found correlations between the gradient of the feature map in the 
orientation variable and the orientation itself(fig. \ref{fig:p:corr}d). 
This means that the orientation  of a receptive field tends to be aligned 
parallel to the direction of the stripe of common orientational 
preference it resides in. This may be due to the overlap of similarly 
oriented neurons being greater when the neurons are aligned in the 
same direction of the orientation, leading to a mutual strengthening of 
the development of those neurons. It may also be indirectly caused by 
horizontal and vertical artefacts occuring in both the grid of LGN 
connections and the rectangular array used for V1.

\subsection{Eye Growth During Critical Period}

It has been found experimentally (in cats) that the lateral extent of ennervation of LGN neurons on the retina is relatively constant, while the diameter of the eye changes dramatically\cite{OLS80,RUSO77}. This causes the 
receptive field size of LGN neurons to decrease by a factor of four 
during development. In the  V1 layer, this translates to larger receptive 
fields and a greater overlap in neighboring receptive fields during 
development as compared to in the mature system. We found it 
necesary to include this effect in our model in order to obtain good 
orientational feature maps.

In previous theories, the question of the extent of receptive field 
overlap was overlooked. One either had complete overlap, or the 
network was so small the overlap was effectively complete. Such a 
situation allows one to use algorithms such as Kohonen's Algorithm,:
\begin{equation}\label{eq:koho} 
\Delta \bf{w}_{r} = h_{rr'}(\bf{y} - \bf{w}_{r})
\end{equation} 
where $r$ is the index of the neuron being updated, $r'$ is the index of 
the neuron with maximal response to the input, and $h_{rr'}$ 
summarizes the lateral interaction, usually as some kind of gaussian. In 
this case each neuron recieves the same input, $\bf{y}$. This does not 
occur in the visual systems under study, in which receptive fields may 
have considerable overlap, but do not overlap completely. The $\bf{y}$ 
in eq. \ref{eq:koho} would have to be replaced by a $\bf{y}_r$, denoting 
the input into the receptive field of the specific neuron $r$. If 
$\bf{y}_r$ changes too quickly as a function of $r$, the algorithm of the 
present model does not perform well. In our simulations, we found that using a 
receptive field size for the fully-grown eye to result in maps that 
typically contained only a few singular points and had broad regions of 
near-constant orientational preference. This experience seems to 
indicate that eye growth is not only accounted for primary visual cortex 
development, but plays an important role.

\section{The Neural Microcircuit Suggested by the Model}

Parvocellular neurons of the LGN terminate in Layer 4A and 
Layer 4C$\beta$ of the primary visual cortex. Magnocellular neurons 
terminate in layer 4B$\alpha$. The neurons in these layers have small 
receptive fields and are primarily nonoriented, though some orientation 
is found in 4B$\alpha$. The receptive fields are approximately the same 
as those of the LGN neurons from which the signal comes. Therefore the 
nonoriented cells of Layer 4C$\beta$ are the first layer  of the proposed 
circuit. The second layer must also be in Layer 4C$\beta$. Oriented cells 
of large receptive field are found in layer 5, and so  this must be the 
next layer of the circuit. These then feed to layer 6, which feeds back to 
the LGN. 

The model would be unchanged if the layer 6 neurons inhibit the 
neurons of Layer 4C$\beta$ instead of the geniculate neurons. One 
simply replaces the interpretations of lgn activities with Layer 
4C$\beta$ activities everywhere. This is possible, however, it does not 
explain the purpose of the dense connections from layer 6 to the LGN 
and does not provide for the quenching of LGN activity.

Orientationally  selective cells in layer 4C$\alpha$ might arise from 
connections between Layer 4C$\beta$ and Layer 4C$\alpha$. Since the magnocellular path seems to be constructed for vision under low-light 
circumstances, the summation of weak Layer 4C$\beta$ activities into 
the 4C$\alpha$ layer could enhance the signal there. Then there would 
be an opportunity for orientational selectivity to form as described 
above. 

\begin{figure}[p]
\vspace{-2.8cm}
\begin{center}
\begin{tabular}[t]{c}
\subfigure[]{\parbox{\textwidth}{
\epsfxsize=\textwidth
\epsfysize=0.45\textwidth
\epsffile{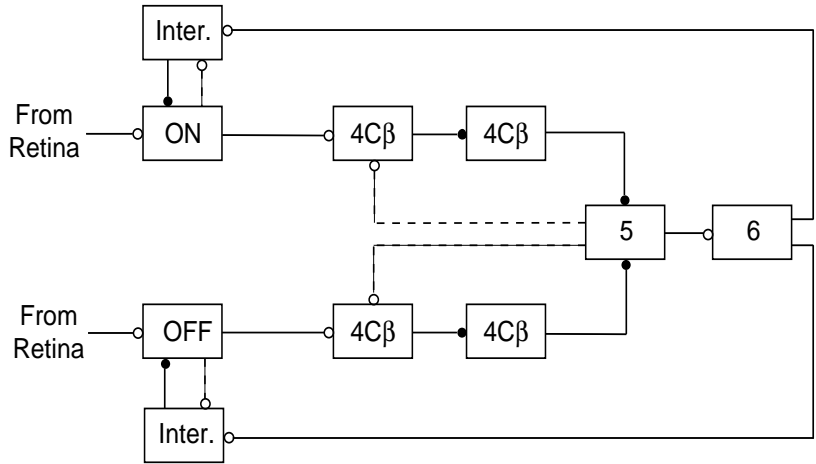}
}
}
\vspace{-0.0cm}\\
\subfigure[]{\parbox{\textwidth}{
\epsfxsize=\textwidth
\epsfysize=0.45\textwidth
\epsffile{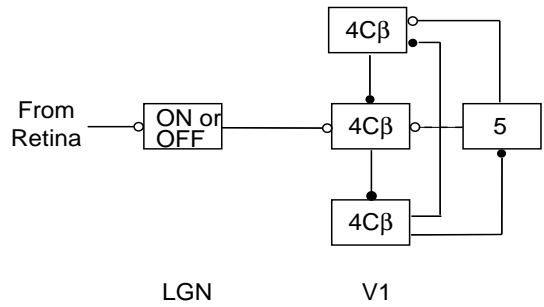}
\vspace{-0.15cm}
}
}
\end{tabular}
\end{center}
\vspace{-0.5cm}
\caption[]{{Functional Microcircuits for LGN-V1 interaction.} a)
Proposed connection scheme between one V1 orientation column and one LGN cell. Open circles are excitatory connections, closed circes are inhibitory. Each V1 orientation column is connected to many LGN cells through replicas of this circuit. The circuit is the same for either on-center or off-center geniculate cells. Sequence of inhibitory interneurons provides for exciation upon favorable input, inhibition upon unfavorable. Dashed lines serve to correlate activities of neurons they connect, and are active only during learning stage.
b) Alternative Circuit when the backwards inhibition occurs within the V1 layer.}

\label{fig:p:microcircuit}
\end{figure}

The algorithms proposed in this paper for the development of 
orientational feature maps may be implemented by a connection 
scheme illustrated in figure \ref{fig:p:microcircuit}. This shows the 
connections between an orientation column in V1 and an LGN neuron. 
Each V1 orientationally-selective neuron is connected to many ON-
center and OFF-center LGN neurons through replicas of this circuit. The 
pattern of the ON-center and off-center LGN neuron is such as to form a 
feature detector, as described previously. The microcircuit is the same 
for ON-center or OFF-center LGN neurons. Other microcircuits for the 
visual cortex have been proposed in refs.\cite{DOUG91D,SILL85A}.

It is known that connections from the LGN to V1 are excitatory. 
However, rather than connect the geniculate cell directly to the 
orientationally-selective V1 neuron through an excitatory connection, 
we connect it through two inhibitory interneurons in sequence. This 
allows a bimodal response for the orientationally-selective V1 neuron. 
When the geniculate neuron receives a favorable input(e.g., light for 
an ON-center cell, and dark for an OFF-center cell) it excites the first 
inhibitory interneuron, which inhibits the inhibition of the second  
interneuron  on the V1 cell. This results in an effective excitation of the 
V1 cell. If the geniculate neuron receives an unfavorable input(e.g., 
dark for an ON-center cell, and light for an OFF-center cell)it excites the 
first inhibitory interneuron less(we assume that there is some level of 
"idling" activity in all neurons) which causes less inhibition on the 
second inhibitory interneuron, resulting in more inhibition on the 
orientationally-selective V1 neuron. The geniculate neuron is inhibited 
by the orientation column through a layer 6 efferent connection to an 
inhibitory interneuron in the LGN. The interpretation of negative 
activities as being the amount by which the activity is below the idling 
value is consistent with this circuit. 

The hebbian update rule used in the algorithm of the simulation was 
based on correlation between the activity of the LGN neuron and the 
orientationally-selective V1 neuron. Since the hebbian-learned 
connection is between the LGN neuron and the first inhibitory 
interneuron, to be consistent with the hebbian rule we add an 
excitatory connection from the orientationally-selective V1 neuron to 
the first inhibitory interneuron. This correlates the activities of the 
orientationally-selective V1 neuron and the first inhibitory neuron. 
Likewise, an excitatory connection from the LGN neuron to the LGN 
inhibitory interneuron correlates the activities for the learning of the 
backwards connections. These connections are illustrated by dashed 
lines in the figure to show that they are active specifically during the 
learning stage.

How do updates occur for negative activity values? If negative activity 
values are interpreted as activity levels below the idling activity, then 
the hebbian learning would be less, not more. One hypothesis is that the 
connection strengths naturally decay over time, and that the idling 
activity is such as to cause learning which more or less balances the 
decay. Then activities below the idling value cause learning less than 
required to overcome the decay and hence the connection strength 
decreases. Such a decay is also supposed to keep the connection 
strengths from eventually all saturating. A reduction of the plasticity to 
zero after the learning stage can also accomplish this, too.

\section{Discussion}

\subsection{The Effect of Retinal Preprocessing}

The preprocessing by filtering through the center-surround receptive fields of the retina is important for the function of the network described here. The removal of the broad intensity components of the original  input images allows the V1 receptive fields to develop into edge detectors. Otherwise the receptive fields would simply detect the overall intensity levels in the image. We limited the network to learning only edge detectors. The same algorithm can produce receptive fields with more periods. This would undoubtably improve the representation, since the combining of edge detectors to make higher frequency components would not then be necessary. Such a mixture of detector types is found in the brain\cite{MULL84B}.

\subsection{The Functional Purpose of Inhibitory Feedback}

The effect in the LGN of the inhibitory feedback can be understood in an 
intuitive way by noting that the receptive field of the input neuron in 
layer 4 of V1 is  exactly matched by the field of the output neuron in 
layer 6. Therefore inhibition of the layer 6 neuron takes away just that 
activity in the LGN which caused the activation of the layer 4 input neuron.

The purpose of this circuit is to prevent the same information from  
being  repeatedly sent from the LGN to V1, which would be 
metabolically disadvantageous. Metabolic energy available in the brain 
is a critical quantity. The same mitochondria which power  brain cells 
drive other cells of the body and thus must be conserved for the overall 
benefit of the organism. Additionally, the temperature of the brain must 
be controlled precisely, and the problem here is usually on the side of 
overheating. Therefore it is quite plausible that the brain employs 
methods to process information which minimize the use of metabolic 
energy.

The result of the theory is that even for static images the neural activity 
is not static; one would expect a burst of activity upon the presentation 
of a new image, followed by a quick dampening. Simulations of this 
effect and comparisons to experiment are shown later. 

Another advantage of such processing comes about when a part of the 
visual scene is moving with respect to a background. If the background 
is stationary, the resulting V1 activities become low, while a moving 
object will cause high activities because the dampening effect is not 
immediate. Thus the brain responds best to changes in the visual scene, 
which again saves metabolic energy for those aspects which are most 
important to be perceived, and at the same time highlights them. 

Furthermore, image features which cannot be captured by the receptive 
fields of  V1 neurons cannot be cancelled by them, either, and therefore 
remain in the LGN  activities. In the present context, this is known as 
Gibbs phenomenon. The result of this is that such features might then 
be learned through the hebbian rule strengthening the connections 
between the LGN  and V1 based on the high activity of this LGN pattern. 
Therefore there is a way in which a natural hierarchy of feature 
detectors in V1 may be learned. 

\subsection{History of Concepts Used in this Theory}

The idea  that orientational selectivity is the result of an alignment of 
afferent LGN cells terminating onto a V1 neuron is due to Hubel and 
Wiesel\cite{HUBE62}, who also pioneered much of the 
neurophysiological work in the visual system. Recently, the work of 
Chapman, et al. \cite{CHAP91A},has shown that each V1 neuron 
typically has many afferent LGN relay cells, and for some of them 
the LGN cells are arranged in a line.  Our model agrees with the above 
hypothesis, in that the arrangement of afferent LGN cells forms the 
receptive field of the V1 neuron. Rather than modelling this LGN cluster 
as a line, however, in order to model edge-detecting cells rather than 
cells sensitive to a bar, we find that the LGN cells connected to an edge-
detecting neuron in V1 are arranged as a group of on-center cells on 
one side of the receptive field, and a group of off-center cells on the 
other side. This approach is found in most of the work cited in the 
introduction.  In the model presented herein, we show that this 
arrangement results from hebbian learning between the LGN and V1 
neurons. In the appendix, we show that the hebbian rule results from 
maximizing the rate of  information transfer from the LGN to V1.

The orientational feature map is less frequently modeled, especially by 
the direct application of lateral interactions. To our knowledge, only von 
der Malsburg applied all the interactions explicitly.We acknowledge the 
pre-existing models of Linsker and Miller, et al, and for purposes of 
clarity would like to describe some differences between the present 
work and theirs. 

Both authors used an ensemble averaging approach. That is, the inputs 
to the model were expected correlation functions of the raw inputs to 
the algorithms. The differential equations describing the updating of the 
weight values was then solved. In the present model, all connections are 
modeled explicitly.

The present model also employs inhibitory feedback. We have outlined 
the functional advantages of the feedback. In the technical aspects of 
the simulation, the feedback also has the effect of controlling the growth 
of the weight values. In all other high-dimensional models of cortical 
development, it is necesary to continously renormalize or otherwise 
decrease the weight values in order to keep them from taking on large 
values. In the present model, should the weight values become large, 
the inhibition acts to decrease LGN activity all the more, resulting in a 
decrease of weight magnitudes. 

The decay process is ultimately responsible for any reduction in weight 
strength. We assume that learning associated with nominal activity is 
just enough to balance out the decay rate. Normally, the concern with 
such an assumption is that it is unstable; should the spontaneous 
activity fall below the required amount, the decay of the connection 
would cause the activity to become even less, and if the spontaneous 
activity becomes too high, the opposite effect occurs. However, in the 
present model, the hebbian learning will bring the connection  strengths 
up to their correct value in the former case, and the inhibitory feedback 
will modulate the growth of the connection strength in the latter case. 
Thus, besides its utility in the efficiency of information transfer, 
inhibition may in general play a role in the maintenance of connection  
strengths in the presence of hebbian learning with decay. 

We have used the receptive fields of LGN neurons to construct the 
receptive fields of V1 neurons. Thus only sixteen connections between 
the LGN and each V1 neuron are required to for the receptive field. This 
saves an enormous amount of computational time compared to 
employing direct connections to each input pixel, equivalent to using 
input receptive fields of one pixel in size.

Thus our model could be made large enough to actually process images 
of appreciable size. In a mature network, with the weight updates 
turned off, a 304 by 304 input image can  be presented to the retina 
and the subsequent activities in the LGN and V1 can be calculated 
according to the equations in the description of the model. In this way, 
the dynamics of cortical image processing can be investigated. 
Results show that the time series of V1 activities obtained 
bear qualitative resemblance to those obtained from experiment. This investigation is described in the next chapter.

Other authors have incorporated recurrent connections into neural 
network models \cite{GROS76A,GROS80A,CARP89A,TAVA90A,OKAJ91A,HAUS90A,KAMI93}, although usually in an enhancement mode for the sake of 
pattern recognition, the recurrent connections being excitatory rather 
than inhibitory.  These are models of pattern recognition in which 
stored patterns feedback into the input layer so as to reinforce 
matching patterns in the input. We do not necessarily disagree with these models, except that we think  that if it occurs, it probably occurs in higher brain areas than those described in our study, areas more directly involved in pattern recognition. But once a pattern is recognized, we feel that the activities which caused it could well be inhibited. Even V1 or V2 activities may be inhibited by backward connections from the inferotemporal cortex upon recognition of the pattern that caused them.

This process of recognition results in  a locus of activity that moves 
from the input sensory receptors to progressively higher areas of brain 
function. In the end, the consciousness is fixated on the thoughts caused 
by a sensory input, rather than the  input itself.

Some authors have speculated on the role of backward connections. 
Pece\cite{PECE92A} proposed inhibitory connections from V1 to the 
LGN. Plumbley described a similar relation within V1. Harth has 
proposed a processing scheme whereby the LGN is used as a center in 
which the input images are modified, perhaps enhanced or irrelevant 
features subtracted as  a result of processing by the cortex. This point of 
view is very close to the theory presented here. However in our theory, 
the features represented in the LGN are not enhanced, but rather 
diminished as  the information about them is transferred to V1. This 
prevents the same information  from being sent repeatedly to V1, and 
enables the activity in V1 to be focussed on aspects of the image not yet 
processed, either in the case of some part of the is moving while the 
rest remains stationary, or in the case in which some part of the image 
is more complex than the rest and requires more time to process.
Mumford\cite{MUMF91A,MUMF92A} suggested that with the backward 
connections from V1 and other areas,the LGN  served as a sort of 
"blackboard" to keep track of the processing already done by the cortex. 
In this view, feedback from several areas of the cortex converge on the 
LGN, and the results are compared to form a coherent understanding of 
the input. 

\subsection{Conclusion}

We have shown that a linear network with lateral connections suffices 
to produce orientational feature maps that compare well with 
experimentally-obtained maps. We found that the filtering of the input 
image through the center-surround receptive fields of retinal-ganglion 
cells is crucial to the formation of orientated antagonistic excitatory and 
inhibitory receptive fields in V1. Eye growth during development is an 
important factor in the development of the feature map, as more 
highly-correlated inputs are needed for the algorithms to be effective. 
Feedback from V1 to the LGN plays the role of modulating growth in the 
connection strengths and  preventing redundant information from being 
transferred from the LGN to V1. As a result, all signals in the LGN and 
V1 are transient, even for stationary images. The model presented here 
is capable of calculating time series for LGN and V1 activities for natural 
images presented to the retina. With the learning turned off, a mature 
network becomes a working image processing system under the other 
equations presented herein, which models the working of the visual 
system of the brain. Simulations of many of neurophysiological 
experiments have been carried out and are described in the next chapter.

\section{Deriving the Hebbian Learning Rule}\label{sec:hebbrule}

Let us symbolize matrices and vectors by uppercase and lowercase bold 
roman letters, respectively. Define  {\bf y} as a vector with components $y_j$ running over the index $j$, similarly {\bf a} as $a_k$ over the components $k$, and  ${\bf w}_k$ as $w_{jk}$ over the index $j$. Let ${\bf W}$ be $w_{jk}$ as a matrix; it is formed by taking  the vectors ${\bf w}_k$  as its rows. Then equation  \ref{eq:v1act} is
\begin{equation}\label{eq:v1actmat}
{\bf a} = {\bf H W y}
\end{equation}
With $ V  =  \bf{H W}$, one can define the signal at the V1 layer by
\begin{eqnarray}\label{eq:p:7.5}
S &=& {\bf a \cdot a}\\
\nonumber\\
  &=& {\bf V y \cdot a}\\
\nonumber
\end{eqnarray}
and derive the learning rule by maximizing the signal by gradient 
ascent, 
\begin{eqnarray}\label{eq:deltv1}
\Delta  {\bf V}_{jl} &=& \frac{\partial S}{\partial {\bf V}_{jl}}\\
\nonumber\\
&=&  y_j a_{l}\\
\nonumber
\end{eqnarray}
Now, 
\begin{eqnarray}\label{eq:deltv}
\Delta {\bf V} &=&  {\bf H} \: \Delta {\bf W}
\end{eqnarray}
since $\bf H$ is constant. If $\bf H=I$, then the hebbian rule is obtained immediately. 
\begin{eqnarray}\label{eq:p:2}
\Delta  w_{jl} &=&  y_j a_{l}
\end{eqnarray}
In any case, $\bf H$ from the simulations is 
invertable, so multiplying \ref{eq:deltv} by ${\bf H}^{-1}$ we obtain
\begin{eqnarray}\label{eq:p:2}
\Delta  w_{jl} &=&  y_j [{\bf H}^{-1} {\bf a}]_{l}
\end{eqnarray} 
In our model, the lateral interaction $\bf H$ was applied to the V1 
activities $\bf a$ only one time, but a closer approximation to reality 
would be to apply $\bf H$ many times, until a fixed point is reached. 
Then,
\begin{eqnarray}\label{eq:p:2}
\bf H^{-1} {\bf a} = {\bf a}
\end{eqnarray}
and the hebbian rule is again derived. Maximizing the signal is 
equivalent to maximizing the information transfer rate
\begin{eqnarray}\label{eq:p:2}
R =  \ln (1+\frac{S}{N})
\end{eqnarray}
where $N$, the noise added to the signal, is a nonzero constant.

\eject
\begin{figure}[p]
\epsfxsize=\hsize
\epsffile{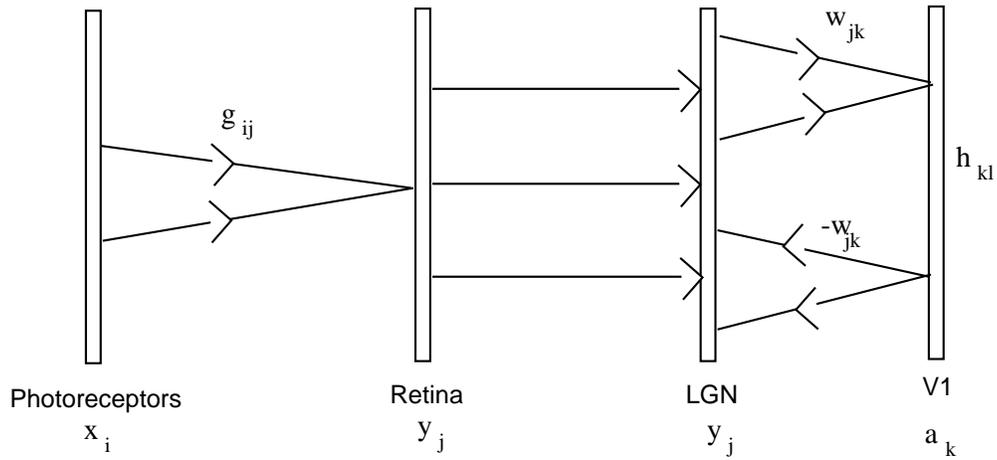}
\caption[]{ Neural network used in the simulation. The network uses feedforward connections from the retinal photoreceptors to the retinal ganglion cells of the center-surround type, one-to-one feedforward connections from the retinal ganglion cells to neurons in the lateral geniculate nucleus, feedforward connections from the lateral geniculate nucleus to the primary visual cortex which develop into oriented edge detectors, and feedback connections from the primary visual cortex to the lateral geniculate nucleus which are of the same form as the feedforward connections. The variables used for the activities of cells in the various layers are shown below the corresponding layer.}
\label{fig:p:network}
\end{figure}

\bibliographystyle{unsrt}
\bibliography{diss}

\begin{thebibliography}{10}

\bibitem{MOUN78A}
V.B. Mountcastle.
\newblock An organizing principle for cerebral function: the unit module and
  the distributed system.
\newblock In G.~M. Edelman and V.B. Mountcastle, editors, {\em The Mindful
  Brain}, pages 7--50. MIT Press, 1978.

\bibitem{BLAS86}
G.G. Blasdel and G.~Salama.
\newblock Voltage sensitive dyes reveal a modular organization in monkey
  striate cortex.
\newblock {\em Nature}, 321:579--585, 1986.

\bibitem{RITT91A}
Helge Ritter, Thomas Martinetz, and Klaus Schulten.
\newblock {\em Textbook: Neural Computation and Self-Organizing Maps: An
  Introduction}.
\newblock Addison-Wesley, New York, revised english edition, 1992.

\bibitem{HUBE62}
D.~H. Hubel and T.~N. Wiesel.
\newblock Receptive fields, binocular interaction, and functional architecture
  in the cat's visual cortex.
\newblock {\em J. Physiol.}, 160:106--154, 1962.

\bibitem{HUBE74B}
D.~H. Hubel and T.~N. Wiesel.
\newblock Uniformity of monkey striate cortex: A parallel relationship between
  field size, scatter, and magnification factor.
\newblock {\em J. Comp. Neurol.}, 158:295--306, 1974.

\bibitem{MALS73}
C.~von~der Malsburg.
\newblock Self-organization of orientation sensitive cells in the striate
  cortex.
\newblock {\em Kybernetik}, 14:85--100, 1973.

\bibitem{BLAS91B}
G.~Blasdel.
\newblock Orientation selectivity, preference and continuity in monkey striate
  cortex.
\newblock {\em J. Neurosci.}, 12:3139--3161, 1992.

\bibitem{CHAP91A}
B.~Chapman, K.~R. Zahs, and M.~P. Stryker.
\newblock Relation of cortical cell orientation selectivity to alignment of
  receptive fields of the geniculocortical afferents that arborize within a
  single orientation column in ferret visual cortex.
\newblock {\em J. Neurosci.}, 11:1347--1358, 1991.

\bibitem{PECE92A}
A.~E.~C. Pece.
\newblock Redundancy reduction of a gabor representation: A possible
  computational role for feedback from primary visual cortex to lateral
  geniculate nucleus.
\newblock In I.~Aleksander and J.~Taylor, editors, {\em Artificial Neural
  Networks, 2: proceedings of the 1992 International Conference on Artificial
  Neural Networks (ICANN-92), Brighton, United Kingdom, 4-7 September, 1992},
  pages 865--868. North-Holland, Amsterdam, 1992.

\bibitem{DAUG85A}
J.~G. Daugman.
\newblock Representational issues and local filter models of two-dimensional
  spatial visual encoding.
\newblock In D.~Rose and V~G Dobson, editors, {\em Models of the Visual
  Cortex}, pages 96--107. John Wiley and Sons, Ltd., London, 1985.

\bibitem{SANG89}
T.~Sanger.
\newblock Optimal unsupervised learning in a single-layer feedforward neural
  network.
\newblock {\em Neural Networks}, 2:459--473, 1989.

\bibitem{YUIL89A}
D.~S.~Cohen A.~L.~Yuille, D. M.~Kammen.
\newblock Quadrature and the development of orientation selective cortical
  cells by hebb rules.
\newblock {\em Biol. Cybern.}, 61:183--194, 1989.

\bibitem{KAMM88}
D.~M. Kammen and A.~L. Yuille.
\newblock Spontaneous symmetry-breaking energy functions and the emergence of
  orientation selective cortical cells.
\newblock {\em Biol. Cybern.}, 59:23--31, 1988.

\bibitem{HANC92}
P.~Hancock, R.~Baddeley, and L.~Smith.
\newblock The principal components of natural images.
\newblock {\em Network}, 3:61--70, 1992.

\bibitem{WORG90}
F.~W\"org\"otter, E.~Niebur, and C.~Koch.
\newblock Modeling visual cortex, hidden anisotropy in an isotropic inhibitory
  connection scheme.
\newblock In R.~Eckmiller, editor, {\em Neuronal Networks for Sensory and Motor
  Systems}. Elsevier, 1990.

\bibitem{WORG92B}
Florentin W\"org\"otter, Ernst Niebur, and Christof Koch.
\newblock Generation of direction selectivity by isotropic intracortical
  connections.
\newblock {\em Neural Computation}, 4:332--340, 1992.

\bibitem{WORG93}
F.~Worgotter and E.~Niebur.
\newblock Cortical column design: a link between the maps of preferred
  orientation and orientation tuning strength?
\newblock {\em Biological Cybernetics}, 70:1--13, 1993.

\bibitem{HUMM79}
R.~Hummel.
\newblock Feature detection using basis functions.
\newblock {\em Computer Graphics and Image Processing}, 9:40--55, 1979.

\bibitem{MARR82A}
D.~Marr.
\newblock {\em Vision}.
\newblock W. H. Freeman, San Francisco, 1982.

\bibitem{RUBN90}
Jeanette Rubner and Klaus Schulten.
\newblock Development of feature detectors by self-organization: A network
  model.
\newblock {\em Biol. Cybernetics}, 62:193--199, 1990.

\bibitem{RUBN90A}
Jeanette Rubner, Klaus Schulten, and Paul Tavan.
\newblock A self-organizing network for complete feature extraction.
\newblock In {\em Proceedings of the Int. Conf. on Parallel Processing in
  Neural Systems and Computers {(ICNC)}, D\"usseldorf}, pages 365--368,
  Amsterdam, 1990. Elsevier.

\bibitem{KAMI93}
R.~Kamimura.
\newblock Internal representation with minimum entropy in recurrent neural
  networks: Minimizing entropy through inhibitory connections.
\newblock {\em Network}, 4:423--440, 1993.

\bibitem{LIZH93}
Z.~Li and J.~J. Atick.
\newblock Toward a theory of the striate cortex.
\newblock {\em Neural Computation}, 6:127--146, 1994.

\bibitem{NASS75}
M.~M. Nass and L.~N. Cooper.
\newblock A theory for the development of feature detecting cells in visual
  cortex.
\newblock {\em Bilogical Cybernetics}, 19:1--18, 1975.

\bibitem{COOP79}
L.~N. Cooper, F.~Liberman, and E.~Oja.
\newblock A theory for the acquisition and loss of neuron specificity in visual
  cortex.
\newblock {\em Biological Cybernetics}, 33:9--28, 1979.

\bibitem{WATS90}
A.~B. Watson.
\newblock Algotecture of visual cortex.
\newblock In C~Blakemore, editor, {\em Vision:coding and efficiency}, pages
  393--410. Cambridge Universtiy Press, Cambridge, 1990.

\bibitem{MCIL91}
W.~McIlhagga.
\newblock A model for simple cells as optimal edge detectors.
\newblock {\em Biol. Cybern.}, 66:177--183, 1991.

\bibitem{ROJE90}
A.~S. Rojer and E.~L. Schwartz.
\newblock Cat and monkey cortical columnar patterns modeled by
  bandpass-filtered 2d white noise.
\newblock {\em Biol. Cybern.}, 62:381--391, 1990.

\bibitem{ROJE90A}
A.~S. Rojer and E.~L. Schwartz.
\newblock Cat and monkey cortical columnar patterns modeled by
  bandpass-filtered 2d white noise.
\newblock {\em Biol. Cybern.}, 62:381--391, 1990.

\bibitem{SWIN82}
N.~V. Swindale.
\newblock A model for the formation of orientation columns.
\newblock {\em Proc. R. Soc. Lond.}, 215:211--230, 1982.

\bibitem{SWIN91}
N.~V. Swindale.
\newblock Coverage and the design of the striate cortex.
\newblock {\em Biol. Cybern.}, 65:415--424, 1991.

\bibitem{SWIN92}
N.~V. Swindale.
\newblock A model for the coordinated development of columnar systems in
  primate striate cortex.
\newblock {\em Biol. Cybern.}, 66:217--230, 1992.

\bibitem{DURB90}
R.~Durbin and G.~Mitchinson.
\newblock A dimension reduction framework for understanding cortical maps.
\newblock {\em Nature}, 343:644--647, 1990.

\bibitem{GOOD90A}
G.~J. Goodhill and D.~J. Willshaw.
\newblock Application of the elastic net algorithm to the formation of occular
  dominance stripes.
\newblock {\em Network}, 1:41--59, 1990.

\bibitem{TAKE79A}
A.~Takeuchi and S.~Amari.
\newblock Formation of topographic maps and columnar microstructures in nerve
  fields.
\newblock {\em Biol. Cybern.}, 35:63--72, 1979.

\bibitem{OBER90A}
Klaus Obermayer, Helge Ritter, and Klaus Schulten.
\newblock A principle for the formation of the spatial structure of cortical
  feature maps.
\newblock {\em Proc. Natl. Acad. Sci. USA}, 87:8345--8350, 1990.

\bibitem{OBER90C}
Klaus Obermayer, Helge Ritter, and Klaus Schulten.
\newblock A neural network model for the formation of topographic maps in the
  {CNS}: Development of receptive fields.
\newblock In {\em International Joint Conference on Neural Networks, San Diego,
  California}, volume~2, pages 423--429. The Institute of Electrical and
  Electronics Engineers, New York, 1990.

\bibitem{OBY-91B}
Klaus Obermayer, Gary~G. Blasdel, and Klaus Schulten.
\newblock A neural network model for the formation and for spatial structure of
  retinotopic maps, orientation- and ocular dominance columns.
\newblock In Teuvo Kohonen, Kai M\"akisara, Olli Simula, and Jari Kangas,
  editors, {\em Artificial Neural Networks}, pages 505--511. Elsevier,
  Amsterdam, 1991.

\bibitem{OBY-92C}
Klaus Obermayer, Klaus Schulten, and Gary Blasdel.
\newblock Statistical-mechanical analysis of self-organization and pattern
  formation during the development of visual maps.
\newblock {\em Phys. Rev. A}, 45(10):7568--7589, May 1992.

\bibitem{MALS77A}
C.~von~der Malsburg and D.~J. Willshaw.
\newblock How to label nerve cells so that they can interconnect in an ordered
  fashion.
\newblock {\em Proc. Natl. Acad. Sci. USA}, 74:5176--5178, 1977.

\bibitem{MALS81}
C.~{von der Malsburg}.
\newblock The correlation theory of brain function.
\newblock Internal Report 81--2, Department of Neurobiology, Max Planck
  Institute for Biophysical Chemistry, G\"ottingen, FRG, 1981.

\bibitem{LINS86}
R.~Linsker.
\newblock From basic network principles to neural architecture: Emergence of
  orientation columns.
\newblock {\em Proc. Natl. Acad. Sci. USA}, 83:8779--8783, 1986.

\bibitem{LINS90A}
R.~Linsker.
\newblock Self-organization in a perceptual system: How network models and
  information theory may shed light on neuronal organization.
\newblock In S~J Hanson and C~R Olson, editors, {\em Connectionist Modelling
  and Brain Function}, pages 351--392. MIT Press, Cambridge, Mass., 1990.

\bibitem{LINS86A}
R.~Linsker.
\newblock From basic network principles to neural architecture: Emergence of
  spatial-opponent cells.
\newblock {\em Proc. Natl. Acad. Sci. USA}, 83:7508--7512, 1986.

\bibitem{MILL89}
K.~D. Miller, J.~B. Keller, and M.~P. Stryker.
\newblock Ocular dominance column development: Analysis and simulation.
\newblock {\em Science}, 245:605--615, 1989.

\bibitem{MACK90}
D.~J.~C. Mackay and K.~D. Miller.
\newblock Analysis of linsker's simulations of hebbian rules.
\newblock {\em Neural Computation}, 2:173--187, 1990.

\bibitem{MILL94}
K.~D. Miller.
\newblock A model for the development of simple cell receptive fields and the
  ordered arrangement of orientation columns through activity-dependent
  competition between on- and off-center inputs.
\newblock {\em J. of Neurosci.}, 14:409--441, 1994.

\bibitem{OBER91}
Klaus Obermayer, Helge Ritter, and Klaus Schulten.
\newblock A model for the development of the spatial structure of retinotopic
  maps and orientation columns.
\newblock {\em {IEICE} Trans. Fund. Electr. Comm. Comp. Sci.},
  {E75-A}(5):537--545, May 1992.

\bibitem{ERWI94}
E.~Erwin, K.~Obermeyer, and K.~Schulten.
\newblock Models of orientation and ocular dominance columns in the visual
  cortex: A critical comparison(submitted).

\bibitem{KOHO82B}
T.~Kohonen.
\newblock Self-organized formation of topologically correct feature maps.
\newblock {\em Biol.~Cybern.}, 43:59--69, 1982.

\bibitem{HEBB49A}
D.~O. Hebb.
\newblock The first stage of perception.
\newblock In D~O Hebb, editor, {\em The Organization of Behavior}, pages
  xi--xix, 60--78. Wiley, New York, 1949.

\bibitem{ORBA84}
G.A. Orban.
\newblock {\em Neuronal Operations in the Visual Cortex}.
\newblock Springer-Verlag, New York, 1984.

\bibitem{WIES74A}
T.~N. Wiesel and D.~H. Hubel.
\newblock Ordered arrangement of orientation columns in monkeys lacking visual
  experience.
\newblock {\em J. Comp. Neurol.}, 158:307--318, 1974.

\bibitem{OLS80}
C.~R. Olson and R.~D. Freeman.
\newblock Rescaling of the retinal map of visual space during growth of the
  kitten's eye.
\newblock {\em Brain Research}, 186:55--65, 1980.

\bibitem{RUSO77}
A.~C. Rusoff and M.~W. Dubin.
\newblock Development of receptive-field properties of retinal ganglion cells
  in kittens.
\newblock {\em J. of Neurophys.}, 40:1188--1198, 1977.

\bibitem{DOUG91D}
R.~J. Douglas and K~.A.~C. Martin.
\newblock A functional microcircuit for the cat visual cortex.
\newblock {\em J. of Physiology}, 440:735--769, 1991.

\bibitem{SILL85A}
A.~M. Sillito.
\newblock Inhibitory circuits and orientation selectivity in the visual cortex.
\newblock In D.~Rose and V~G Dobson, editors, {\em Models of the Visual
  Cortex}, pages 96--107. John Wiley and Sons, Ltd., London, 1985.

\bibitem{MULL84B}
W.~H. Mullikin, J.~P. Jones, and L.~A. Palmer.
\newblock Periodic simple cells in cat area 17.
\newblock {\em J. of Neurophys.}, 52:372--387, 1984.

\bibitem{GROS76A}
S.~Grossberg.
\newblock Adaptive pattern classification and universal recoding: I. parallel
  development and coding of neural feature detectors.
\newblock {\em Biol. Cybern.}, 23:121--134, 1976.

\bibitem{GROS80A}
S.~Grossberg.
\newblock How does the brain build a cognitive code?
\newblock {\em Psychological Review}, 87:1--51, 1980.

\bibitem{CARP89A}
G.~Carpenter.
\newblock Neural network models for pattern recognition and associative memory.
\newblock {\em Neural Networks}, 2:243--257, 1989.

\bibitem{TAVA90A}
P.~Tavan and H.~Kuhnel.
\newblock Self-organization of associative memory and pattern classification:
  recurrent signal processing on topological feature maps.
\newblock {\em Biol. Cybern.}, 64:95--105, 1990.

\bibitem{OKAJ91A}
K.~Okajima.
\newblock A recurrent system incorporating characteristics of the visual
  system: a model for the function of backward neural connections in the visual
  system.
\newblock {\em Biol. Cybern.}, 65:235--241, 1991.

\bibitem{HAUS90A}
G.~Hausler.
\newblock Chaos, cooperation, and associative memory in nonlinear pictoral
  feedback systems.
\newblock In R.~Eckmiller, G~Hartmann, and G~Hauske, editors, {\em Parallel
  Processing in Neural Systems and Computers}, pages 533--538. Elsevier Science
  Publishers, 1990.

\bibitem{MUMF91A}
D.~Mumford.
\newblock On the computational architecture of the neocortex i. the role of the
  thalamo-cortical loop.
\newblock {\em Biol. Cybern.}, 65:135--145, 1991.

\bibitem{MUMF92A}
D.~Mumford.
\newblock On the computational architecture of the neocortex ii. the role of
  cortico-cortical loops.
\newblock {\em Biol. Cybern.}, 66:241--251, 1992.

\end{thebibliography}
\eject

\end{document}